\documentclass[reprint,amsmath,amssymb,aip]{revtex4-2}
\usepackage{graphicx}
\usepackage{dcolumn}
\usepackage[usenames,dvipsnames]{color}

\usepackage{bm}
\usepackage{booktabs,dcolumn,caption} 
\usepackage[table]{xcolor} 
\usepackage[normalem]{ulem} 

\def \be {\begin{equation}}
\def \ee {\end{equation}}
\def \ben {\begin{eqnarray}}
\def \een {\end{eqnarray}}
\def \bi {\begin{itemize}}
\def \ei {\end{itemize}}

\usepackage{multirow}
\usepackage{ctable}

\definecolor{dkgreen}{rgb}{0,0.6,0}
\definecolor{gray}{rgb}{0.5,0.5,0.5}
\definecolor{mauve}{rgb}{0.58,0,0.82}
\usepackage{listings}

\begin{document}

\title{Simple and general unitarity conserving numerical real time propagators of time dependent Schr\"{o}dinger equation based on Magnus expansion}
\author{Taner M. Ture}
\affiliation{Department of Chemistry and Biochemistry, Queens College, City University of New York, 65-30 Kissena Boulevard, Queens, New York 11367, USA \& PhD Programs in Chemistry and Physics, Graduate Center of the City University of New York, New York 10016, USA}
\author{Seogjoo J. Jang}
\email{seogjoo.jang@qc.cuny.edu}
\affiliation{Department of Chemistry and Biochemistry, Queens College, City University of New York, 65-30 Kissena Boulevard, Queens, New York 11367, USA \& PhD Programs in Chemistry and Physics, Graduate Center of the City University of New York, New York 10016, USA}

\date{Published in the {\it Journal of Physical Chemistry A} {\bf 128}, 2871-2882 (2024)}

\begin{abstract}
Magnus expansion (ME) provides a general way to expand the real time propagator of a time dependent Hamiltonian within the exponential such that the unitarity is satisfied at any order.  We use this property and explicit integration of Lagrange interpolation formulas for the time dependent Hamiltonian within each time interval and derive approximations that preserve unitarity for the differential time evolution operators of general time dependent Hamiltonians.    The resulting second order approximation is the same as using the average of Hamiltonians for two end points of time.  We identify three fourth order approximations involving commutators of Hamiltonians at different times, and also derive a sixth order expression.   Test of these approximations along with other available expressions for a two state time dependent Hamiltonian with sinusoidal time dependences provides information on relative performance of these approximations, and suggests that the derived expressions can serve as useful numerical tools for time evolution for time resolved spectroscopy, quantum control, quantum sensing, real time {\it ab initio} quantum dynamics, and open system quantum dynamics. 
\end{abstract}

\maketitle
\section{Introduction}
Accurate numerical integration of time dependent Schr\"{o}dinger equation with time dependent Hamiltonian is crucial for reliable modeling of time resolved spectroscopy,\cite{mukamel,cina,haacke-burghardt,jang-exciton} quantum control,\cite{shapiro-qc,bergmann-rmp70,singh-jcp150,hu-jcp152} and more recently  quantum sensing.\cite{degen-rmp89-2017} Computationally, it is also an important issue for real time {\it ab initio} quantum dynamics\cite{castro-jcp121,goings-wcms2017} and for open system quantum dynamics employing quantum master equations\cite{jang-jcp129,jang-jcp131,jang-jcp135,jang-njp15,jang-jcp157,jang-exciton,montoya-castillo-jcp143,mulvihill-jcp156,devega-rmp89}  in the interaction picture.\cite{jang-jcp129,jang-jcp131,jang-jcp135,jang-njp15,jang-jcp157}  
Thus, developing an accurate, efficient, and  general numerical approximation for the real time propagator of a time dependent Hamiltonian has broad implications and applications.  One key property that needs to be ensured for such approximation is the conservation of unitarity.  If this is not satisfied, unphysical behavior including violation of norm conservation is expected to occur.  Methods based on the Trotter\cite{schulman,trotter-pamc-10} and Suzuki factorizations\cite{suzuki-jmp26,suzuki-jmp38} or their generalizations\cite{jang-jcp115,chin-pla226} have played important roles in this respect, but their extension to time dependent Hamiltonian remain difficult  in general although some progress was made.\cite{chin-jcp117}  Time ordered exponential operator serves as a general formal solution, but its simple perturbative truncation  is not guaranteed to preserve unitarity.   Magnus expansion (ME)\cite{magnus-cpam7,pechukas-jcp44,burum-prb24,prato-jcp106,blanes-pr470} serves as a formally superior starting point in this respect because it is designed to preserve unitarity at any finite order of series and is applicable to any time dependent Hamiltonian.  However, the actual use of ME based propagators has been rather limited to specialized areas of research\cite{evans-ap48,takegoshi-jcp142,singh-jcp150,begzjav-rp17,mananga-pr609,mananga-cpl730} and its tests\cite{milfeld-pra27,blanes-pr470,kormann-jcp128} for general numerical quantum dynamics propagation of time dependent Hamiltonians have been scanty to the best of our knowledge.

In a sense, ME can be viewed as a specific expansion of the time ordered exponential operator.\cite{oteo-jmp41}  Then, each term of the Magnus expansion can be considered as a partial re-summation of simple perturbative terms of certain class up to an infinite order.     Indeed, during past couple of decades, significant progress\cite{blanes-pr470} has been made clarifying this view of ME.  A well-defined mathematical criterion for the existence of ME now exists\cite{blanes-pr470,blanes-bnm40} and exact formal expressions up to an infinite order based on the graph theory have been derived.   Most recently, ME  was studied using the theory of graded Lie algebra \cite{munthe-kaas-ptmpes357} and Taylor expansions.\cite{blanes-bnm42} These advances serve as solid basis for employing ME as a practical tool for general quantum dynamics, and have also produced practical algorithms for numerical time evolution.  

In this paper, we show that combining Lagrange interpolation polynomials and ME leads to simple and general expressions for finite order MEs for differential real time propagators, which then can be used for numerical time evolution for a general time dependent Hamiltonian.  We provide detailed derivation of these expressions and also summarize others available,\cite{blanes-bnm40,Iserles-ptmpes357}  which are then tested for a simple two state time dependent model Hamiltonian.   The paper concludes with a summary of major implications and applications of this work.

\section{Magnus expansion up to the fourth order and approximate propagators} 
\subsection{Expressions of ME}
Let us consider a time evolution operator $U(t)$ that satisfies the following time dependent Schr\"{o}dinger equation:
\be
\frac{\partial}{\partial t} U(t)=-\frac{i}{\hbar} H(t) U(t) ,
\ee
where $H(t)$ is an arbitrary time dependent Hamiltonian.  
We are interested in finding an explicit expression for the time evolution operator for $t\in [t_k,t_{k+1}]$ with $t_{k+1}-t_k=\delta t$, which is denoted as $U_{\delta t} (t_{k+1},t_k)$, such that
\be
U(t_{k+1})=U_{\delta t}(t_{k+1},t_{k}) U (t_k) .
\ee 
Application of ME\cite{magnus-cpam7,blanes-pr470}  to $U_{\delta t}(t_{k+1},t_{k})$, employing coefficients slightly different from conventional ones,  leads to the following formal expression:
\be
U_{\delta t}(t_{k+1},t_{k}) =\exp\left \{\sum_{n=1}^\infty \frac{1}{n!} \left (-\frac{i}{\hbar}\right)^n M_n (t_{k+1},t_{k}) \right\} ,  \label{eq:udelta-mn}
\ee
where $M_n(t_{k+1},t_{k})$ up to the fourth order are well known\cite{pechukas-jcp44,blanes-pr470,klarsfield-pra39} and can be expressed as follows: 
\ben 
M_1(t_{k+1},t_k)&=& \int_{t_k}^{t_{k+1}} d\tau H(\tau) , \label{eq:M1} \\
M_2(t_{k+1},t_{k})&=&\int_{t_k}^{t_{k+1}} d\tau \int_{t_k}^{\tau}d\tau' [H(\tau), H(\tau')]  ,\hspace{.3in}  \label{eq:M2}
\een
\ben
M_3(t_{k+1},t_{k})&=&\int_{t_k}^{t_{k+1}} d\tau \int_{t_k}^{\tau} d\tau' \int_{t_k}^{\tau'} d\tau''  \nonumber \\
&&\hspace{.5in} \Big ( \left [ H(\tau),  [H(\tau'),H(\tau'')]\right ]  \nonumber \\
&&\hspace{.5in}+\left [[H(\tau),H(\tau')],H(\tau'') \right ] \Big ) , \label{eq:M3}
\een
\ben
M_4(t_{k+1},t_{k})&=&2\int_{t_k}^{t_{k+1}} d\tau\int_{t_k}^{\tau} d\tau' \int_{t_k}^{\tau'} d\tau''\int_{t_k}^{\tau''} d\tau''' \nonumber \\ 
&&\Big([[[H(\tau),H(\tau')],H(\tau'')],H(\tau''')] \nonumber \\ 
&&+[H(\tau),[[H(\tau'),H(\tau'')],H(\tau''')]] \nonumber \\
&&+[H(\tau),[H(\tau'),[H(\tau''),H(\tau''')]]] \nonumber \\
&&+[H(\tau'),[H(\tau''),[H(\tau'''),H(\tau)]]]\Big) . \label{eq:M4}
\een
Note that expressions for terms of ME beyond the second order are not unique and alternative expressions can be found for the above third and fourth order terms.\cite{blanes-pr470,oteo-jmp41}  

As is clear from the above expressions up to the fourth order, the $n$th order ME term consists of $n$ ordered time integrations.  Recently, a systematic method to generate all the ME terms as ordered integrals of right nested commutators was also developed.\cite{arnal-jpc2}  
This means that unless the time dependent Hamiltonian has singularity, considering up to the $n$th term of ME will ensure that the numerical algorithm is exact up to $\delta t^n$  at the minimum.   In fact, the actual order of accuracy  is even better because the $n$th term involves $n-1$ nested commutators. In addition, terms of even order of $\delta t$ in general disappear once explicit expansion is made with respect to $\delta t$ in a symmetric manner, as shown in Appendix A, which renders the practical order of accuracy of any nominally odd order approximation to become the next order.  More detailed consideration of this is provided below.    

\subsection{Approximations based on polynomial approximations}
We here consider the cases where the time dependent Hamiltonian is smooth enough to allow polynomial expansions within each interval $[t_k,t_{k+1}]$.   More specifically, we assume that $H(t)$ can be approximated as 
\be
H(t)=H^{(l)}(t)+O(\delta t^{l+1}) , 
\ee
where  explicit expressions for $H^{(l)}$ for $l=0,\cdots, 3$ are shown below
\ben
H^{(0)}(t)&=&\frac{1}{2} (H_k+H_{k+1} ) , \\
H^{(1)}(t)&=&H_k \frac{(t_{k+1}-t)}{\delta t}+H_{k+1} \frac{(t-t_k)}{\delta t} , \hspace{.4in}\label{eq:h1t}
\een
\ben
H^{(2)}(t)&=&2H_k\frac{(t_{k+1}-t)(t_{k+\frac{1}{2}}-t)}{\delta t^2}  \nonumber \\
&+&4H_{k+\frac{1}{2}}\frac{(t_{k+1}-t)(t-t_k)}{\delta t^2} \nonumber \\
&+&2H_{k+1}\frac{(t-t_{k+\frac{1}{2}})(t-t_k)}{\delta t^2}  ,  \hspace{.7in}\label{eq:h2t}
\een
\ben
H^{(3)}(t)&=&\frac{9}{2}H_k\frac{(t_{k+1}-t)(t_{k+\frac{2}{3}}-t)(t_{k+\frac{1}{3}}-t) }{\delta t^3}  \nonumber \\
&+&\frac{27}{2}H_{k+\frac{1}{3}}\frac{(t_{k+1}-t)(t_{k+\frac{2}{3}}-t)(t-t_k)}{\delta t^3} \nonumber \\
&+&\frac{27}{2}H_{k+\frac{2}{3}}\frac{(t_{k+1}-t)(t-t_{k+\frac{1}{3}})(t-t_k)}{\delta t^3} \nonumber \\
&+&\frac{9}{2}H_{k+1}\frac{(t-t_{k+\frac{2}{3}}) (t-t_{k+\frac{1}{3}})(t-t_k)}{\delta t^3} .  \label{eq:h3t}
\een
In above expressions, $H_k$ is a short notation for $H(t_k)$.  
We combine the above expressions with MEs and obtain approximations for $U_{\delta t}(t_{k+1},t_k)$ of different order.
To represent the resulting expressions in a compact manner, we here introduce a short notation $M_n^{(l)}$ defined as follows:
\begin{eqnarray}
M^{(l)}_n&\equiv& M_n(t_{k+1},t_{k}) \mbox{ that uses } H^{(l)}(t) \nonumber \\
&&\mbox{ instead of } H(t)\mbox{ in the integrands}\nonumber
\end{eqnarray}

\subsubsection{Second order propagator}
Let us first start with the simplest second order propagator $U_{\delta t}^{(2)}(t_{k+1},t_{k})$ such that 
\be
U_{\delta t}(t_{k+1},t_{k})=U_{\delta t}^{(2)}(t_{k+1},t_k)+O(\delta t^3) .
\ee
To obtain the above second order expression, consideration up to the second order ME is sufficient.  Since each integration over the interval of $\delta t$ increases the order of $\delta t$, we also find that the first and zeroth order approximations for the Hamiltonian are sufficient for the first and the second ME terms respectively as follows: 
\ben
U_{\delta t}^{(2)}(t_{k+1},t_k)=\exp\left \{ -\frac{i}{\hbar} M_1^{(1)} -\frac{1}{2\hbar^2} M_2^{(0)}\right\} .
\een
In the above expression, it is easy to find that 
\be
M_1^{(1)}=\int_{t_k}^{t_{k+1}} d\tau H^{(1)}(\tau)=\frac{\delta t}{2} \left(H_{k}+H_{k+1}\right) .
\ee
On the other hand, $M_2^{(0)}=0$ since $[H^{(0)}(t),H^{(0)}(t)]=0$.   Therefore,
\ben
U_{\delta t}^{(2)}(t_{k+1},t_k)=\exp\left \{ -\frac{i\delta t}{2\hbar} (H_k+H_{k+1} )\right\} .
\een
Thus, the second order approximation for the differential propagator based on ME is nothing but the usual symmetric average approximation of the time integration within the exponent for each propagator.

\subsubsection{Third order propagator}

The third order propagator  $U_{\delta t}^{(3)}(t_k,t_{k+1})$ is nominally accurate up to the order of $\delta t^3$ as follows:
\be
U_{\delta t}(t_{k+1},t_k)=U_{\delta t}^{(3)}(t_{k+1},t_k)+O(\delta t^4) .
\ee
However, in practice, as is shown in Appendix A, $U_{\delta t}^{(3)}(t_{k+1},t_k)$ is in general exact up to the order of $\delta t^4$ since there are no terms of even order with respect to $\delta t$ once all the coefficients are expressed in a symmetrical manner.   
Following the same reasoning as in the second order approximation and noting that $M_3^{(0)}=0$, we find that 
\ben
U_{\delta t}^{(3)}(t_{k+1},t_k)=\exp\left \{ -\frac{i}{\hbar} M_1^{(2)} -\frac{1}{2\hbar^2} M_2^{(1)} \right\}  . \label{eq:u3}
\een
The  two terms within the exponential  in the above expression can be calculated easily, as described in Appendix B, and are 
\ben
&&M_1^{(2)}=\frac{\delta t}{6} (H_k+4H_{k+\frac{1}{2}}+H_{k+1} ) , \label{eq:m1-2}\\
&&M_2^{(1)}=\frac{\delta t^2}{6}[H_{k+1},H_k]  .
\een
In eq \ref{eq:m1-2}, it is important to note that $M_1^{(2)}$ is in fact exact up to the order of $\delta t^4$, which can be proven directly by integration of any cubic function of $t$.  This is because of the cancellation of the cubic term around the quadratic polynomial approximation.  Thus, $M_1^{(2)}$ can still be used for the fourth order propagator being considered  next. 

\subsubsection{Fourth order propagator}
The fourth order propagator  $U_{\delta t}^{(4)}(t_{k+1},t_{k})$ is accurate up to the order of $\delta t^4$ as follows:
\be
U_{\delta t}(t_{k+1},t_k)=U_{\delta t}^{(4)}(t_{k+1},t_k)+O(\delta t^5) .
\ee
Following the same reasoning as in the second and third order approximations and noting that $M_4^{(0)}=0$, we find that 
\ben
U_{\delta t}^{(4)}(t_{k+1},t_k)=\exp\left \{ -\frac{i}{\hbar} M_1^{(2)} -\frac{1}{2\hbar^2} M_2^{(2)} +\frac{i}{6\hbar^3}M_3^{(1)}\right\}  , \nonumber \\ \label{eq:u4-1}
\een
where $M_1^{(2)}$ is given by eq \ref{eq:m1-2} and was used since this is exact up to the order of $\delta t^4$.  $M_2^{(2)}$ and $M_3^{(1)}$ are expressed as  
\ben
&&M_2^{(2)}=\frac{\delta t^2}{30}\left ([H_{k+1},H_k]+4[H_{k+\frac{1}{2}},H_k]+4[H_{k+1},H_{k+\frac{1}{2}}]\right)   \nonumber \\
&&\hspace{.35in}=\frac{\delta t^2}{30}[H_k+4H_{k+\frac{1}{2}},H_k-H_{k+1}] ,\label{eq:m2-2} \\
&&M_3^{(1)}= \frac{\delta t^3}{40}[H_{k+1}-H_k,[H_{k+1},H_k]] . \label{eq:m3-1}
\een
Detailed derivations of the above expressions are also provided in Appendix B. The second expression for $M_2^{(2)}$ in eq \ref{eq:m2-2} is computationally a little more convenient, whereas the first expression makes it clear that $M_2^{(2)}$ is an odd function of $\delta t$.

Close inspection of eq \ref{eq:m3-1} shows that the actual leading order term of $M_3^{(1)}$ is  $O(\delta t^5)$ within the polynomial approximation because $H_{k+1}-H_k=O(\delta t )$.   Therefore, the following approximation is also exact up to the fourth order of $\delta t$:
\ben
U_{\delta t}^{(4)}(t_{k+1},t_k)\approx \exp\left \{ -\frac{i}{\hbar} M_1^{(2)} -\frac{1}{2\hbar^2} M_2^{(2)}\right\}  . \label{eq:u4-2}
\een
Thus, overall, we find that the three approximations, eqs \ref{eq:u3}, \ref{eq:u4-1}, and \ref{eq:u4-2}, are all practically exact up to $\delta t^4$.   
Actual performance of these expressions will be tested through model calculations provided in the next section.

\subsubsection{Fifth order propagator}
The fifth order propagator  $U_{\delta t}^{(5)}(t_k,t_{k+1})$ is nominally accurate up to the order of $\delta t^5$ as follows:
\be
U_{\delta t}(t_{k+1},t_k)=U_{\delta t}^{(5)}(t_{k+1},t_k)+O(\delta t^6) .
\ee
However, this is practically exact up to the order of $\delta t^6$ since there are no even order terms of ME when expanded with respect to $\delta t$ as shown in Appendix A.  

Following the same reasoning as in the second and third order approximations and noting that $M_5^{(0)}=0$, we find that 
\ben
U_{\delta t}^{(5)}(t_{k+1},t_{k})&=&\exp\left \{ -\frac{i}{\hbar} M_1^{(4)} -\frac{1}{2\hbar^2} M_2^{(3)} +\frac{i}{6\hbar^3}M_3^{(2)} \right . \nonumber \\
&&\left . \hspace{.5in} +\frac{1}{24\hbar^3}M_4^{(1)}\right\}   , \label{eq:u6}
\een
where calculation of each term is straightforward.   Final expressions of the first three terms are as follows:
	\ben
		&& M_1^{(4)} = \frac{\delta t}{90}\left(7H_{k}+32H_{k+\frac{1}{4}}+12H_{k+\frac{1}{2}}\right.  \hspace{.7in}\nonumber \\
		&& \hspace{.7in}               \left . +32H_{k+\frac{3}{4}}+7H_{k+1}\right)  , 
\een
\begin{widetext}
\ben
		&& M_2^{(3)} = \frac{\delta t^2}{3360}\left(117([H_{k+\frac{1}{3}},H_k] +[H_{k+1},H_{k+\frac{2}{3}}]) \right . \nonumber \\
		&&  \hspace{.5in}+ 47[H_{k+1},H_{k}] + 144([H_{k+1},H_{k+\frac{1}{3}}] + [H_{k+\frac{1}{3}},H_{k}]) \nonumber \\
		&&\hspace{.5in}\left. + 729[H_{k+\frac{2}{3}},H_{k+\frac{1}{3}}] \right) \nonumber \\
		&&\hspace{.4in} = \frac{\delta t^2}{3360}\left[\frac{232}{39}H_{k}+\frac{1152}{13}H_{k+\frac{1}{3}} + 72H_{k+\frac{2}{3}},2H_{k} + \frac{81}{8}H_{k+\frac{1}{3}} - \frac{13}{8}H_{k+1} \right]\nonumber \\
		&&\hspace{.5in}+\frac{\delta t^2}{90}[H_{k+1},H_{k}] ,  
\een
\end{widetext}
	\ben
		&& M_3^{(2)} =\frac{\delta t^3}{2520}\Big (64[H_{k+\frac{1}{2}}+H_{k+1},[H_{k+\frac{1}{2}},H_{k}]]  \nonumber \\
		&&\hspace{.5in}+64[H_{k+\frac{1}{2}}+H_{k},[H_{k+\frac{1}{2}},H_{k+1}]] \nonumber \\
		&&\hspace{.5in}+44\left([H_{k},[H_k,H_{k+\frac{1}{2}}]]+[H_{k+1},[H_{k+1},H_{k+\frac{1}{2}}]]\right)\nonumber \\
		&&\hspace{.5in}+9[H_{k+1}-H_{k},[H_{k+1},H_{k}]] \Big ) .
         \een
The final term $M_4^{(1)}$ in eq \ref{eq:u6} becomes even more complicated, but we found that it can be simplified as follows:
\be
	 M_4^{(1)} = \frac{\delta t^4}{210}[\frac{1}{c}H_k-H_{k+1},[H_{k+1} - cH_k,[H_{k+1},H_k]]], \label{eq:M_41} 
	\ee
where $c = -\frac{5 - \sqrt{21}}{2}$.  More detailed explanation and its verification using a computer algebra system known as SymPy\cite{meurer-pcs3} are provided  in the Supporting Information (SI).

\subsection{Other approximate expressions\label{other-exp}}
There are other well established approximations\cite{blanes-bnm40,Iserles-ptmpes357} for the differential propagator based on ME, for which we introduce $M(t_{k+1},t_{k})$ such that
	\ben
		U_{\delta t}(t_{k+1},t_{k}) = \exp\left\{-\frac{i}{\hbar} M(t_{k+1},t_{k})\right\} .
		\label{eq:ME time_step}
	\een
Thus, 
\be
M(t_{k+1},t_k)=\sum_{n=1}^\infty \frac{1}{n!} \left (-\frac{i}{\hbar}\right)^{n-1} M_n(t_{k+1},t_k) .
\ee
We here consider four algorithms from other works,\cite{blanes-bnm40,Iserles-ptmpes357} which can be summarized  in our notation as follows:	

\begin{enumerate}
\item {\it 4th order expression using equally spaced points} [Blanes 4th-order]:\cite{blanes-bnm40}
This was derived based on reproducing the Taylor series of $M(t_{k+1},t_{k})$ in terms of univariate integrals, and is expressed in our notation as follows: 
\ben
&&M(t_{k+1},t_{k})= \frac{1}{6}\delta t(H_k+4H_{k+\frac{1}{2}}+H_{k+1}) \nonumber \\
&&- \frac{i}{\hbar}\frac{\delta t^2}{72}[H_{k+1}-H_{k},H_k+4H_{k+\frac{1}{2}}+H_{k+1}] .
\een
Note that this is similar but different from one of our 4th order approximations, eq \ref{eq:u4-2}.
\item {\it 4th order expression using Gauss-Legendre points} [Blanes-4th order (gauss)]:\cite{blanes-bnm40}
This expression was derived using the same univariate integrals as above, but Gauss-Legendre points were used for the integral.  For this, let us define
\ben
&&\gamma_1=\frac{1}{2}-\frac{1}{6}\sqrt{3} , \\
&&\gamma_2=\frac{1}{2}+\frac{1}{6}\sqrt{3} .
\een
Then, employing $H_{k+\gamma_1} = H(t_k+\gamma_1 \delta t)$ and  $H_{k+\gamma_2} = H(t_k+\gamma_2 \delta t)$,  this can be expressed as follows:
\ben
M(t_{k+1},t_{k})&=& \frac{1}{2}\delta t (H_{k+\gamma_1}+H_{k+\gamma_2}) \nonumber \\
&&-\frac{i}{\hbar} \frac{\sqrt{3}}{12}\delta t^2[H_{k+\gamma_2},H_{k+\gamma_1}] .
\een
The above expression can also be obtained using the linear approximation for $M_1$ and $M_2 $ in our expressions through the Gauss-Legendre points.
\item {\it 4th order expression using Gauss-Legendre points (with additional 2nd order commutator)} [Iserles 4th-order (gauss)]:\cite{Iserles-ptmpes357}
This was derived from binary trees by collecting all fourth-order terms and then making polynomial approximation between two points $t_k+\gamma_1\delta t$ and $t_k+\gamma_2\delta t$.  The resulting expression is
	\ben
		&&M(t_{k+1},t_{k})= \frac{1}{2}\delta t(H_{k+\gamma_1}+H_{k+\gamma_2}) \nonumber \\
		&&- \frac{i}{\hbar}\frac{\sqrt{3}}{12}\delta t^2[H_{k+\gamma_2},H_{k+\gamma_1}] \nonumber \\ 
		&&- \frac{1}{\hbar^2}\frac{1}{80}\delta t^3[H_{k+\gamma_2}-H_{k+\gamma_1},[H_{k+\gamma_2},H_{k+\gamma_1}]] .
	\een

\item {\it 6th order expressions using Gauss-Legendre points} [Blanes 6th order (gauss)]:\cite{blanes-bnm40}  
This expression involves evaluation of the Hamiltonian at midpoint $t_{k+\frac{1}{2}}$ and two additional intermediate points.  For this, let us define 
\ben
 &&\alpha_1=\frac{1}{2}-\frac{\sqrt3}{2\sqrt5} , \\
 &&\alpha_2=\frac{1}{2} + \frac{\sqrt3}{2\sqrt5} , 
 \een
and define $H_{k+\alpha_1}=H (t_k+\alpha_1 \delta t)$ and  $H_{k+\alpha_2}= H(t_{k} +\alpha_2 \delta t) $.  
Then, the expression involves first calculating the following operators:
\ben
&&B_0 = \frac{5}{18}(H_{k+\alpha_1}+H_{k+\alpha_2}) + \frac{4}{9}H_{k+\frac{1}{2}} , \\
&&B_1 = \frac{\sqrt{15}}{36}\left (H_{k+\alpha_2}-H_{k+\alpha_1} \right) ,\\
&&B_2 = \frac{1}{24}\left (H_{k+\alpha_1}+H_{k+\alpha_2}\right) .
\een
The above expressions are then used to calculate the following operators:
\ben
&&M_1 = \delta tB_0 , \\ 
&&M_2 = \delta t^2[B_1,3B_0-12B_2] ,  \\
&&\frac{M_3}{6}+\frac{M_4}{24} = \frac{3}{10}\delta t[B_1,M_2] \nonumber \\
&&\hspace{.5in}+  \delta t^2[B_0,[B_0,\frac{\delta t}{2}B_2 - \frac{1}{120}M_2]]  .
\een

\end{enumerate}

It is also useful to provide a general expression based on the Taylor series expansion around $t_{k+\frac{1}{2}}$ here.  
The expressions up to the fourth order ME terms are provided below.\cite{blanes-bnm40} 
\ben
&& M_1 = \delta tc_0 + \delta t^3\frac{1}{12}c_2+\delta t^5\frac{1}{80}c_4 + \mathcal{O}(\delta t^7), \\
&& M_2 = -\delta t^3\frac{1}{6}[c_0,c_1] + \delta t^5\left(-\frac{1}{40}[c_0,c_3] \right . \nonumber \\
&&\hspace{.5in} \left . + \frac{1}{120}[c_1,c_2]\right) + \mathcal{O}(\delta t^7) , \\
&& M_3 = \delta t^5\left(\frac{1}{60}[c_0,[c_0,c_2]] \right . \nonumber \\
&&\hspace{.5in}\left . - \frac{1}{40}[c_1,[c_0,c_1]]\right) + \mathcal{O}(\delta t^7) ,\\
&& M_4 = \delta t^5\frac{1}{30}[c_0,[c_0,[c_0,c_1]]] + \mathcal{O}(\delta t^7) \textcolor{blue}{,}
\een
where $c_n = \frac{1}{n!}\frac{d^nH(t)}{dt^n}|_{t=t_{k+1/2}}$, the $n$th Taylor series coefficient of $H(t)$ at $t=t_{k+1/2}$.

\section{Numerical Test}
For a numerical test of all the approximate expressions for the propagator, we here consider a two state Hamiltonian with sinusoidal time dependences in all three components as follows:  
\ben
H(t)&=&\alpha_1 \sin(\omega_1 t)|1\rangle\langle 1|+(1+\alpha_2\sin (\omega_2 t))|2\rangle\langle 2| \nonumber \\
&&+(1 + \alpha_c \sin(\omega_ct ))(|1\rangle \langle 2|+|2\rangle \langle 1|)  . \label{eq:model-hamil}
\label{eq:Hamiltonian}
\een
The above model represents a physical situation where the two states are coherently modulated by two independent fields with frequencies $\omega_1$ and $\omega_2$ and are also coupled by a third field with frequency $\omega_c$ in addition to an intrinsic coupling.   The choice of parameters in this model amounts to choosing units such that the constant level spacing and coupling between states 1 and 2 are assumed to be one.  In addition, we assume that  $\hbar=1$.  

\begin{figure}	
	(I)\makebox[4.5in]{}\\ \includegraphics[width=0.8\columnwidth]{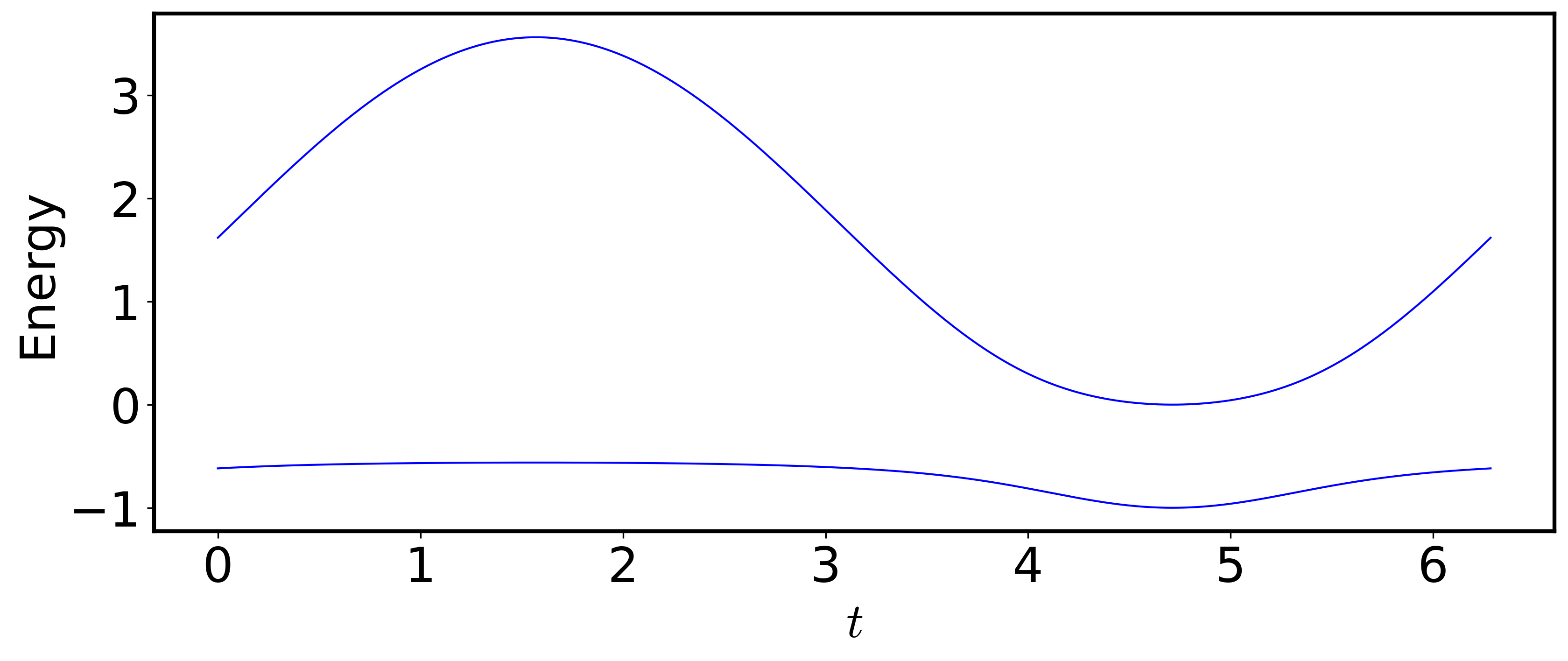} \\
	(II)\makebox[4.5in]{}\\	\includegraphics[width=0.80\columnwidth]{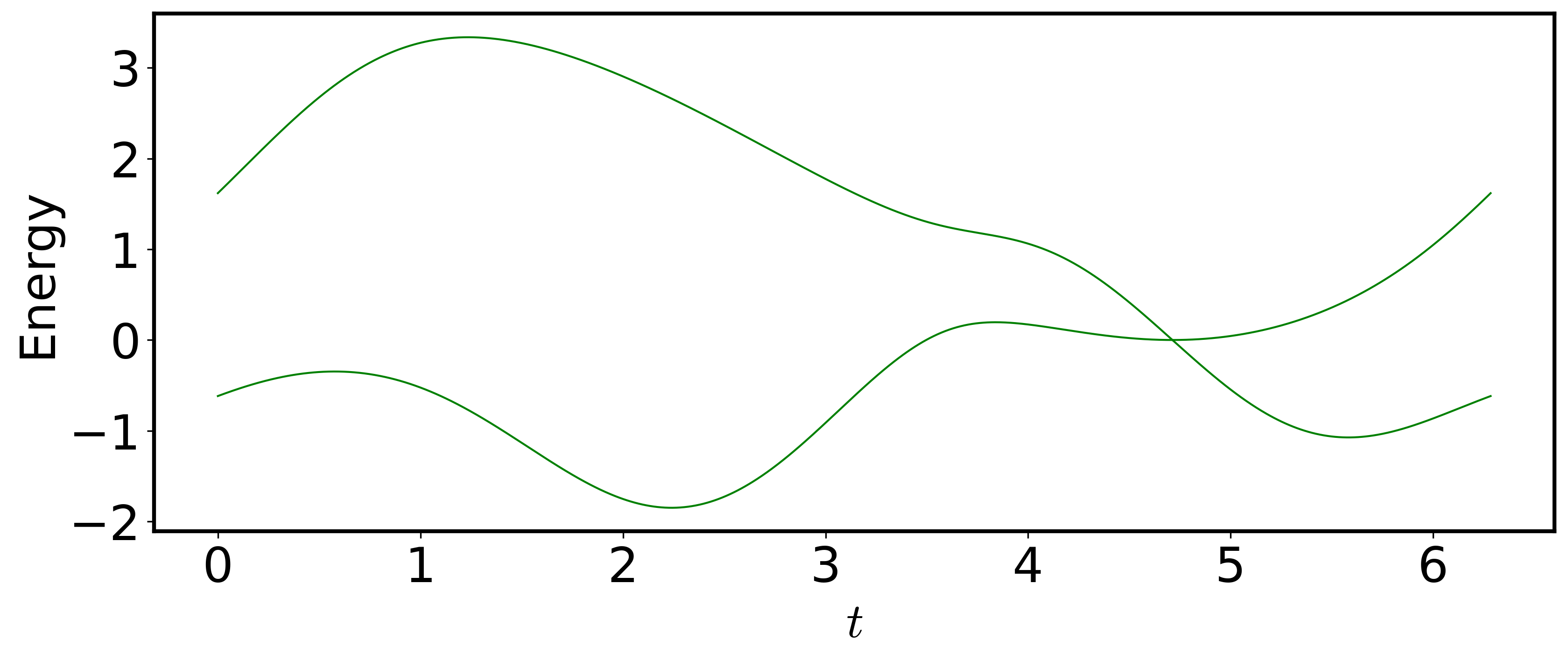} \\
	(III)\makebox[4.5in]{ }\\	\includegraphics[width=0.80\columnwidth]{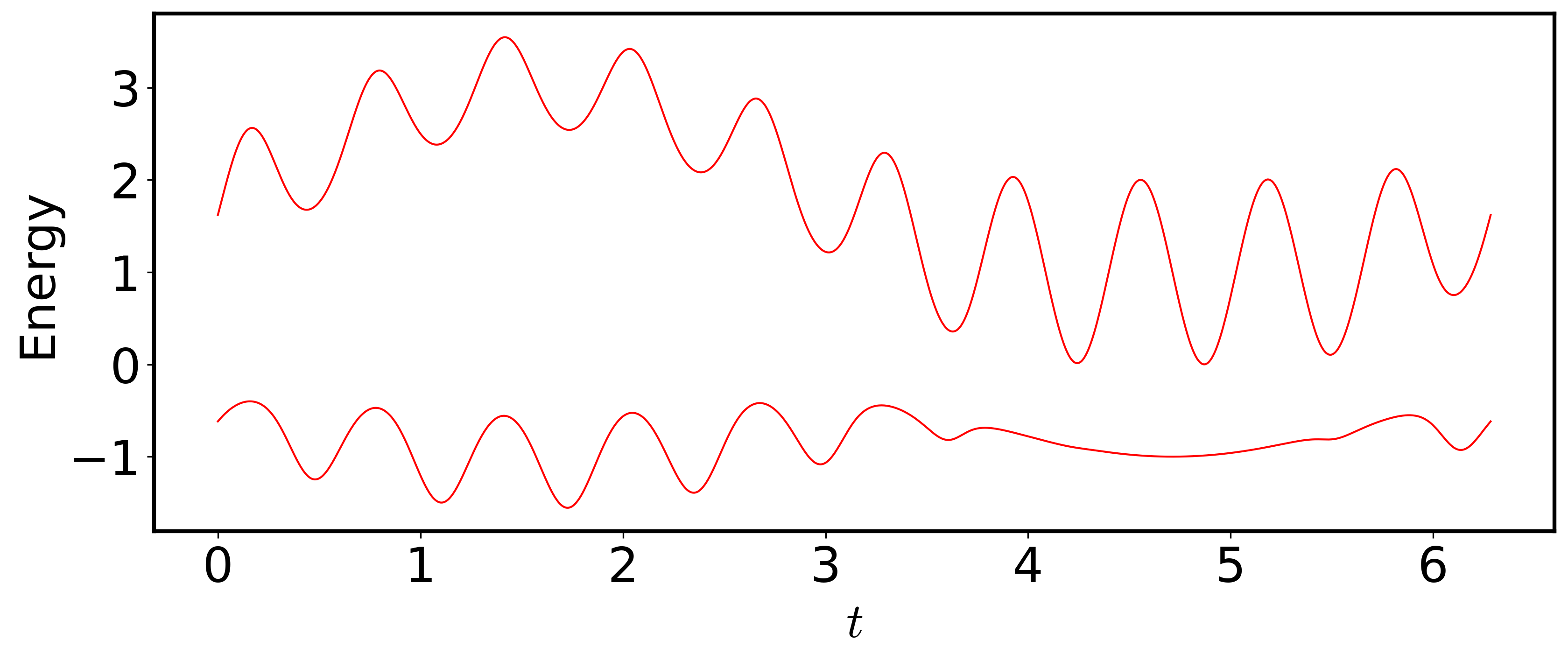} \\
	(IV)\makebox[4.5in]{ }\\	\includegraphics[width=0.80\columnwidth]{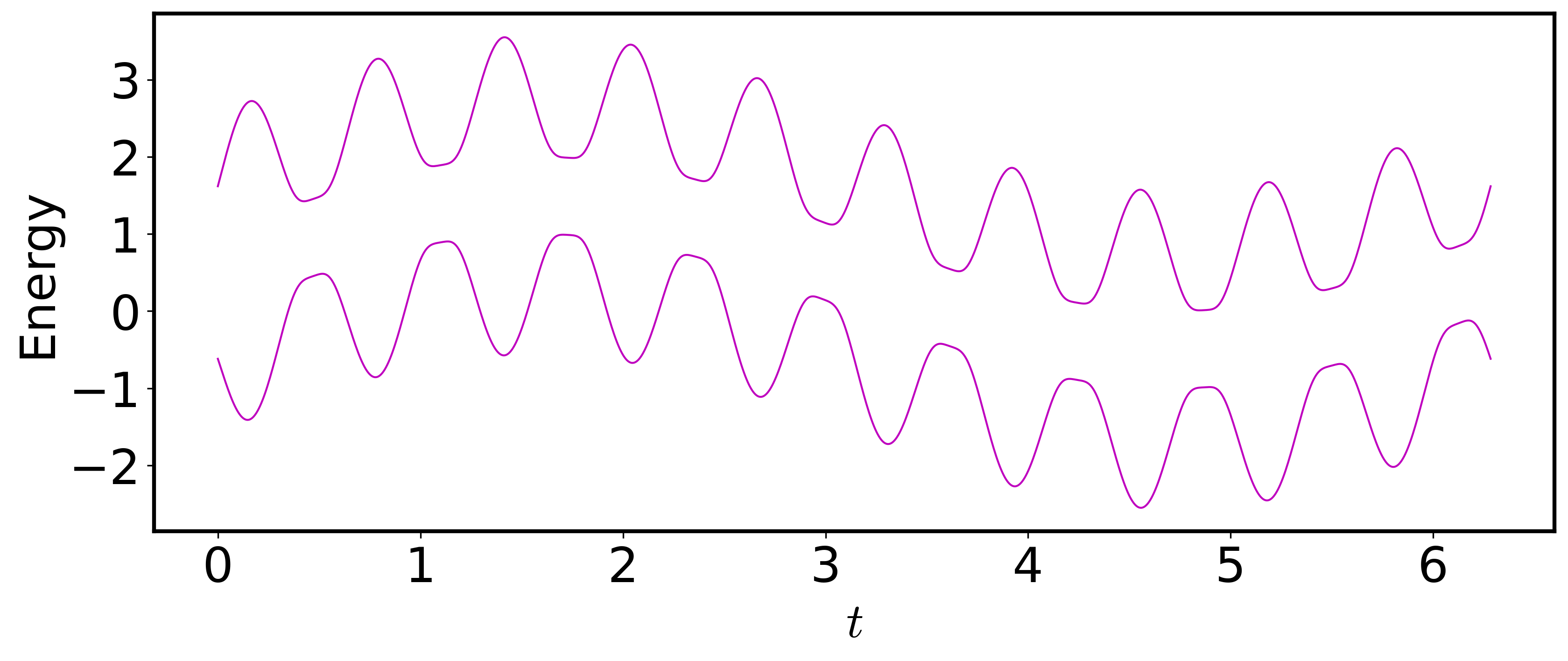} 
	\caption{Time-dependent eigenvalues of  $H(t)$ for the four cases of parameters listed in Table \ref{table-parameter}. The time range $(0,2\pi)$ shown covers a period of the Hamiltonian in all four cases.}
	\label{fig:eigenvalues}
\end{figure}

Note that the main reason for using eq \ref{eq:model-hamil} is not because of its physical significance but rather its utility.  Although simple, the Hamiltonian given by eq \ref{eq:model-hamil} has  non-terminating commutators between those at different values of  time  and is thus appropriate for testing all ME based expressions of different orders of accuracy.   The  sinusoidal time dependences also serve as a more stringent numerical test than polynomial or piecewise polynomial functions of time used in quantum control.  In addition, it is also possible to find analytic expressions for some commutators and simple expressions for  time dependent eigenvalues of the Hamiltonian \cite{hou-fp13} (see Fig. \ref{fig:eigenvalues}), which make it easy to interpret the results of calculation.  Although it is possible to introduce different phase factors within different sine functions, we do not consider such case since it does not appear to be important for the evaluation of numerical propagators.  Even without such phase factors,  the Hamiltonian given by eq \ref{eq:model-hamil} still has 6 parameters.  For the test of approximate ME expressions, we chose four representative cases of parameters  as listed in Table I.

\begin{table}
\caption{Table of parameters for the model Hamiltonian.}
\label{table-parameter}
\begin{tabular}{c|cccccc} \hline \hline
Case &\makebox[.3in]{$\alpha_1$ }&\makebox[.3in]{$\omega_1$} &\makebox[.3in]{$\alpha_2$} &\makebox[.3in]{$\omega_2$}&\makebox[.3in]{$\alpha_c$ }&\makebox[.3in]{ $\omega_c$} \\
\hline
I &1&1&1&1&1&1\\
II&1&2&1&1&1&1\\
III&1&1&1&10&1&1\\
IV&1&1&1&1&1&10\\
\hline
\end{tabular}
\end{table} 

The four parameter sets in Table \ref{table-parameter} cover a broad range of physical and numerical situations.   The case I corresponds to the situation where all time dependences are comparable whereas the case II represents the situation where the modulation of the energy level $1$ is significantly faster than others.  On the other hand, cases III and IV represent two different ways where motional averaging occurs either due to fast modulation of the energy level of state $2$ or the coupling between states $1$ and $2$.  These different situations are well represented by different patterns of time dependent eigenvalues shown in Fig. \ref{fig:eigenvalues}.
The two cases  I and II are quite different from each other because II has a level crossing, for which transformations relying on adiabatic conditions are not possible\cite{jahnke-nm94}. The cases I, III and IV all have only avoided crossings with similar energy gaps, but they differ greatly in the number of avoided crossings and other general features of the time-dependent eigenvalues.
We also  note that, because our numerical methods are based an interpolating polynomial, the higher frequencies of III and IV are more numerically challenging and may require comparatively smaller stepsizes\cite{kormann-jcp128} compared to I and II.   

For each case and each approximate expression, we calculated the propagator $U(t_f,0)$  at $t_f=100$ by repeated application of the approximate expression $U(t_{k+1},t_{k})$, from the left, to $U(t_k,0)$, starting from $U(0,0)={\rm I}$, until $k=N_f=t_f/\delta t$.   All three fourth order expressions we have provided, eqs \ref{eq:u3}, \ref{eq:u4-1}, and \ref{eq:u4-2}, and our sixth order expression, eq \ref{eq:u6}, were tested along with other known fourth and sixth order expressions by Blanes {\it et al.}\cite{blanes-bnm40} and Iserles {\it et al.}\cite{Iserles-ptmpes357} we have summarized.

\begin{figure}		
(I)\makebox[4.5in]{}\\ \includegraphics[width=0.8\columnwidth]{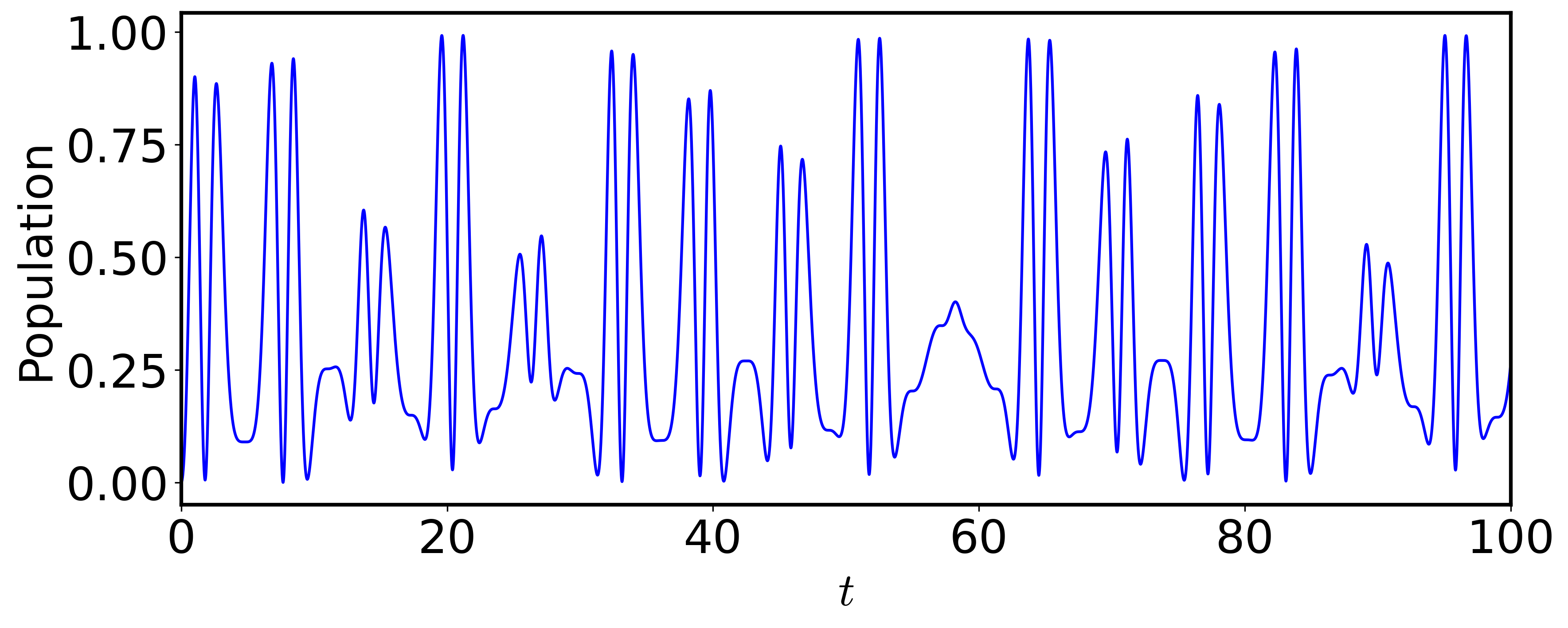} \\
(II)\makebox[4.5in]{}\\	\includegraphics[width=0.8\columnwidth]{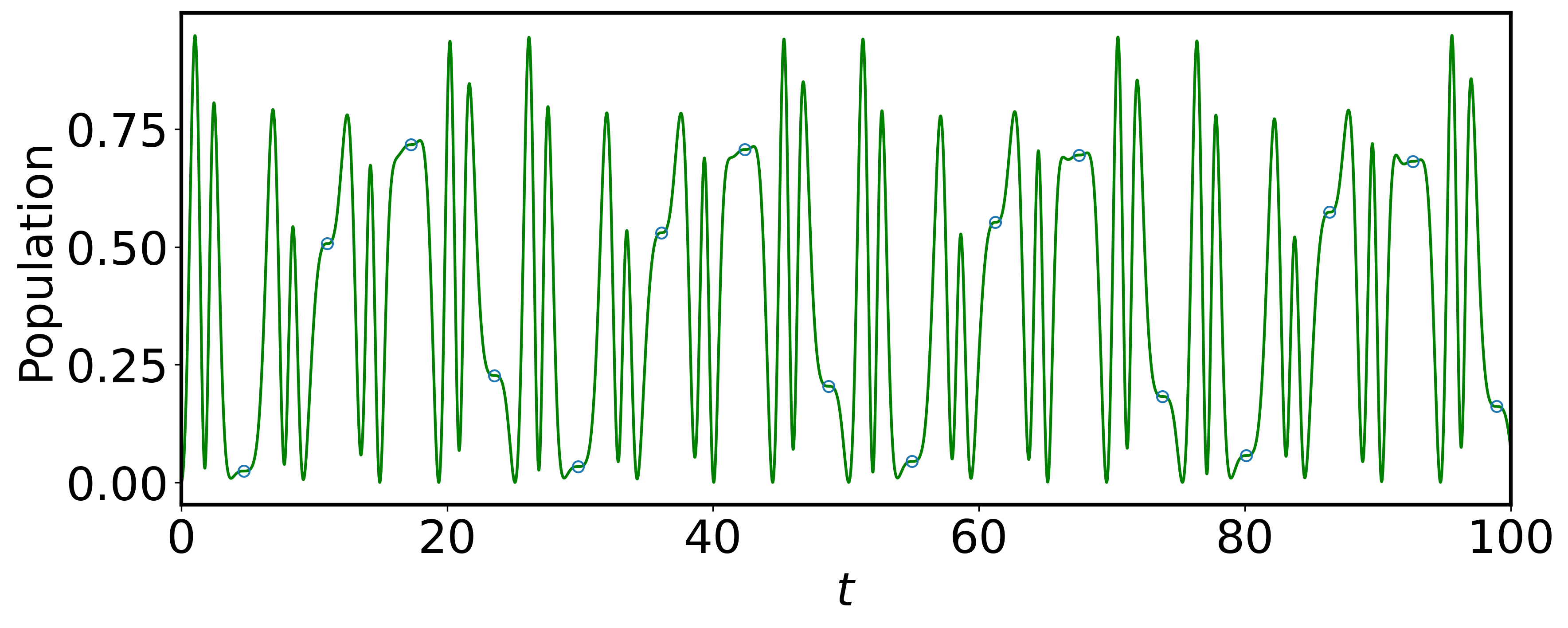} \\
(III)\makebox[4.5in]{ }\\	\includegraphics[width=0.8\columnwidth]{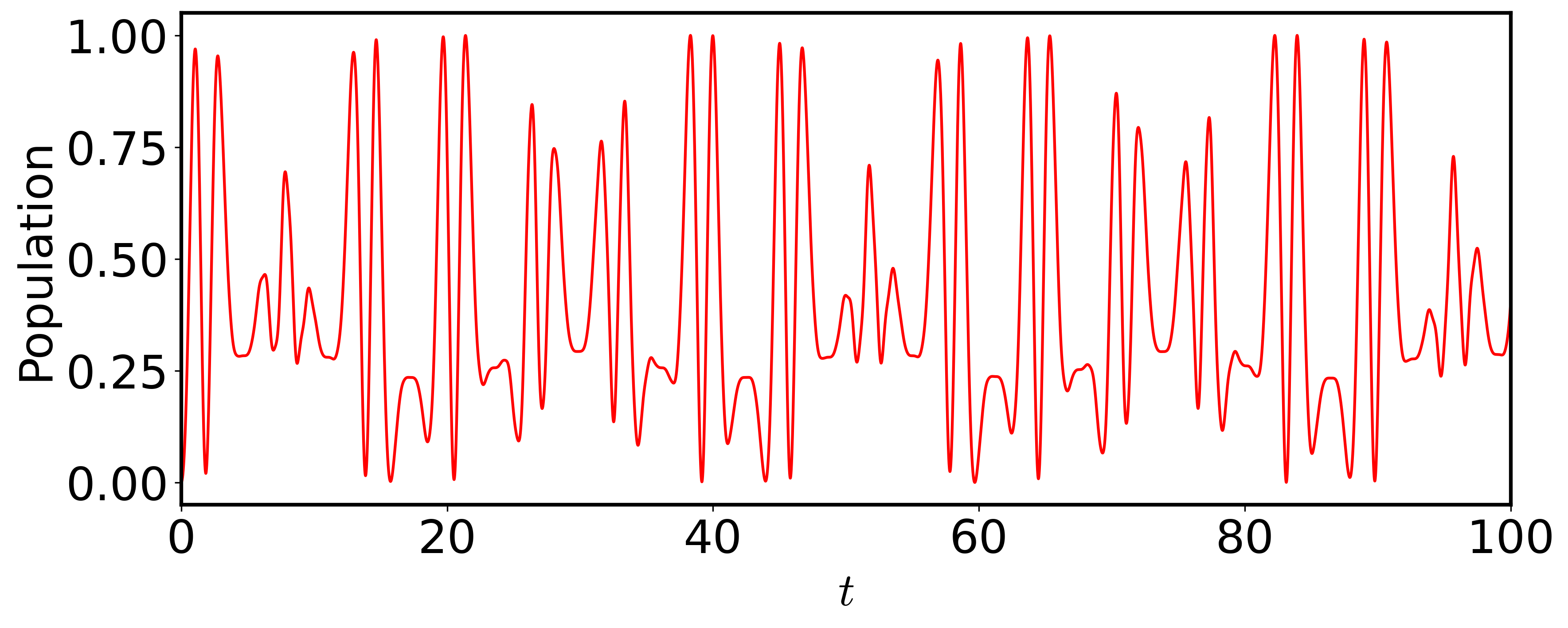} \\
(IV)\makebox[4.5in]{ }\\	\includegraphics[width=0.8\columnwidth]{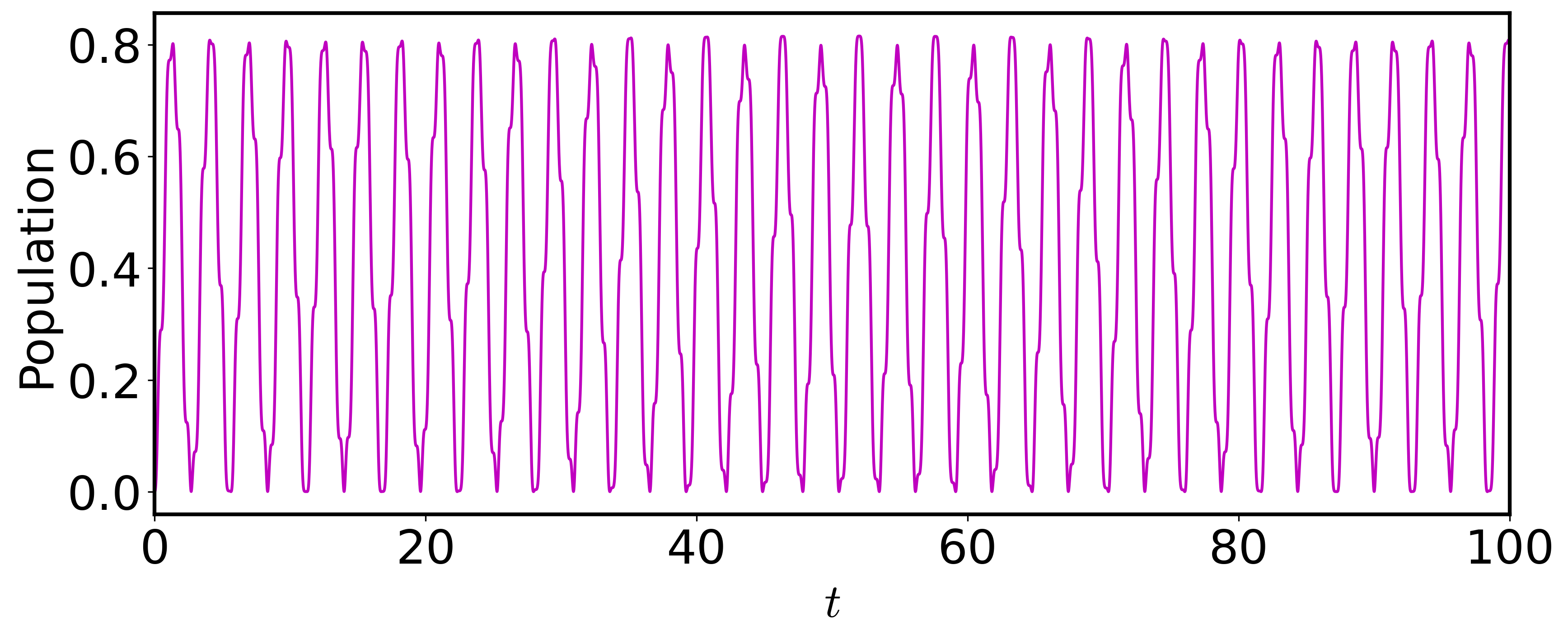} 
	\caption{Populations of  the excited state $|2\rangle$ for the four cases of parameters listed in Table \ref{table-parameter}. At time zero, all the populations are at the state $|1\rangle$. Markers for case II indicate times where level crossings occur, initially at $t \approx 4.712$ and repeating every $2\pi$. }
	\label{fig:populations}
\end{figure}

Figure  \ref{fig:populations} shows time dependent populations at state $|2\rangle$ for the case where the initial state at time $t=0$ is $|1\rangle$. For the calculation of these populations, we used eq \ref{eq:u4-2} for $\delta t=0.00610$, which offers accurate enough results for this demonstration.  The time dependent population for each case shows a different beating structure reflecting the effects of time dependent change of the Hamiltonian components in addition to the intrinsic oscillation due to average off-diagonal coupling.  For the case II, the effect of the level crossing (first occurring at $t \approx 4.712$ and repeating every $2\pi$)  on the populations can be seen. At each level crossing, every vector is an eigenvector of the Hamiltonian.  Thus, no change in population occurs in a small neighborhood around that point. This is why blue markings shown in Fig. \ref{fig:populations} for the case II occur at flat portions of the population where the first derivative is zero.  For the case III, the time dependence of population closely resembles that of the Hamiltonian without the high frequency term (see SI for comparison). Similarly, for the case IV, the population exhibits nearly periodic behavior reminiscent of Rabi oscillations. This can be seen by averaging out the high frequency coupling terms; the resulting Hamiltonian has the same populations as the time-independent one described by $|1\rangle \langle 2| + |2\rangle \langle 1| + |2\rangle \langle 2| $.  As yet, due to the periodic nature of the time dependent Hamiltonian, the populations in all four cases exhibit periodic behavior at sufficiently long time.  It is also clear that all the populations remain within $(0,1)$ as expected from a norm conserving propagator.

 \begin{table}
	\centering
	\caption{Step size and number of points for numerical calculations.}
	\setlength{\tabcolsep}{6pt} 
	\begin{tabular}{c|c}
		\hline
		\hline
		\textbf{Step size} & \textbf{Number of Points} \\
		\hline
		0.00610 & 16384  \\
		0.01221 & 8192  \\
		0.02442 & 4096  \\
		0.04885  & 2048 \\
		
		0.09775& 1024  \\
		0.19569 & 512 \\
		\hline
		\hline
	\end{tabular}
	\label{table:stepsize}
\end{table}
\ \vspace{.4in}

We also calculated errors  in the propagator itself for each case. As a reference for the exact result, we used our sixth order method, eq \ref{eq:u6}, calculated with the choice of $\delta t= 0.00610/8=0.0007625$. Then, for each method and choice of time step, we calculated an error defined as follows:
\be
{\rm error}=\frac{||U_{\rm approx}(t_f)-U_{\rm ref}(t_f)||_F}{||U_{\rm ref}(t_f)||_F}, \nonumber \label{eq:error}
\ee
where the subscript F denotes the Frobenius norm defined by $||U||_F = \left(\sum_i\sum_j |U_{ij}|^2\right)^{\frac{1}{2}}$. The choice of time steps we tested are listed in Table \ref{table:stepsize}. We note that the time steps are in the non-oscillatory ($\omega\delta t \ll 1$ ) and moderately oscillatory regimes ($\omega\delta t = \mathcal{O}(1)$)\cite{Iserlee-ijna24}, where our methods generated from interpolating polynomials are expected to perform well. In addition, even the largest step size used satisfies the convergence criterion\cite{blanes-pr470} of the Magnus Expansion, which is  $\int_{t}^{t+\delta t} \left \|H(\tau)\right\|_2 d\tau < \pi \hbar$, in all four cases. Larger stepsizes are not shown because the data begins to lose its linearity, indicating a loss of convergence. However, the results still have reasonable errors. The fact that truncated propagators can perform well despite large step sizes and divergence of the full ME is an interesting feature of ME propagators that can be explained quantitatively via error bounds. \cite{hochbruck-sjna41}

\begin{figure}
	
	\centering
(I)\makebox[4.5in]{}\\ \includegraphics[width=0.80\columnwidth]{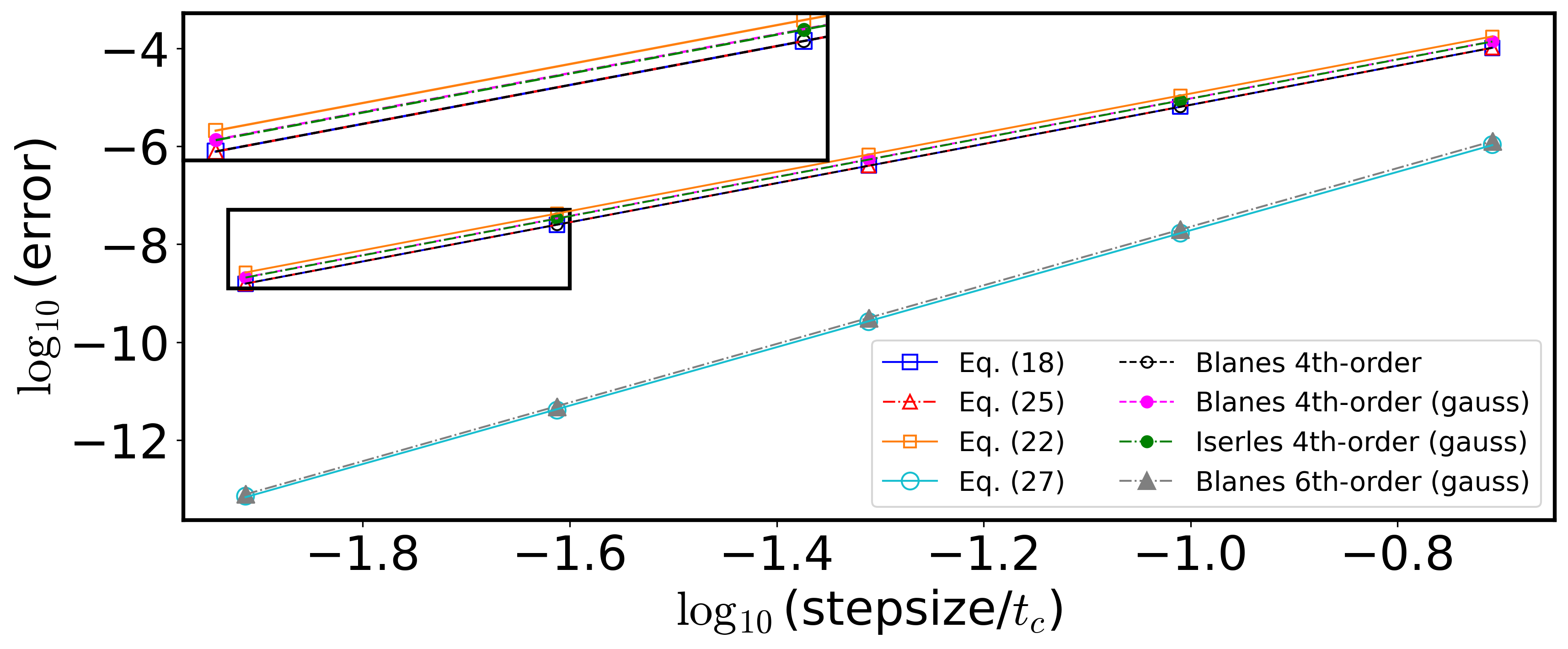} \\
(II)\makebox[4.5in]{}\\	\includegraphics[width=0.80\columnwidth]{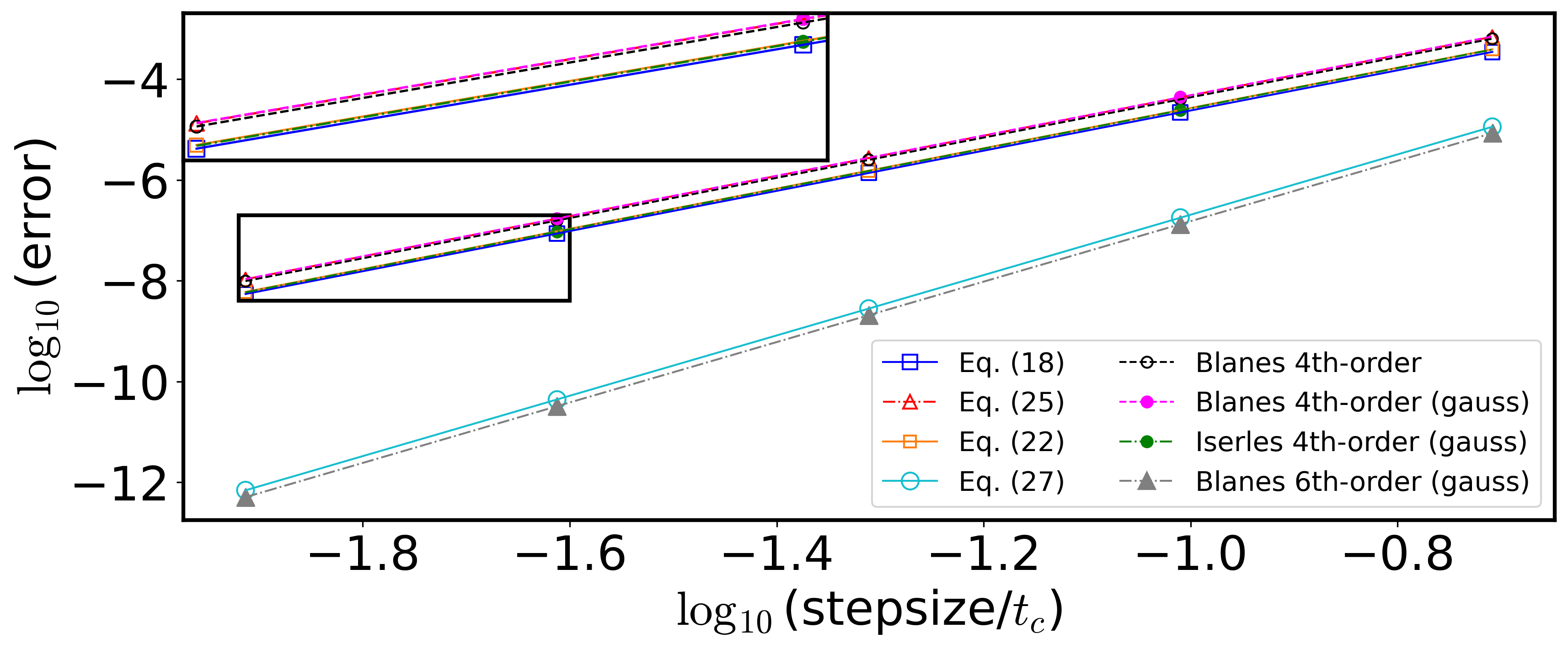} \\
(III)\makebox[4.5in]{ }\\	\includegraphics[width=0.80\columnwidth]{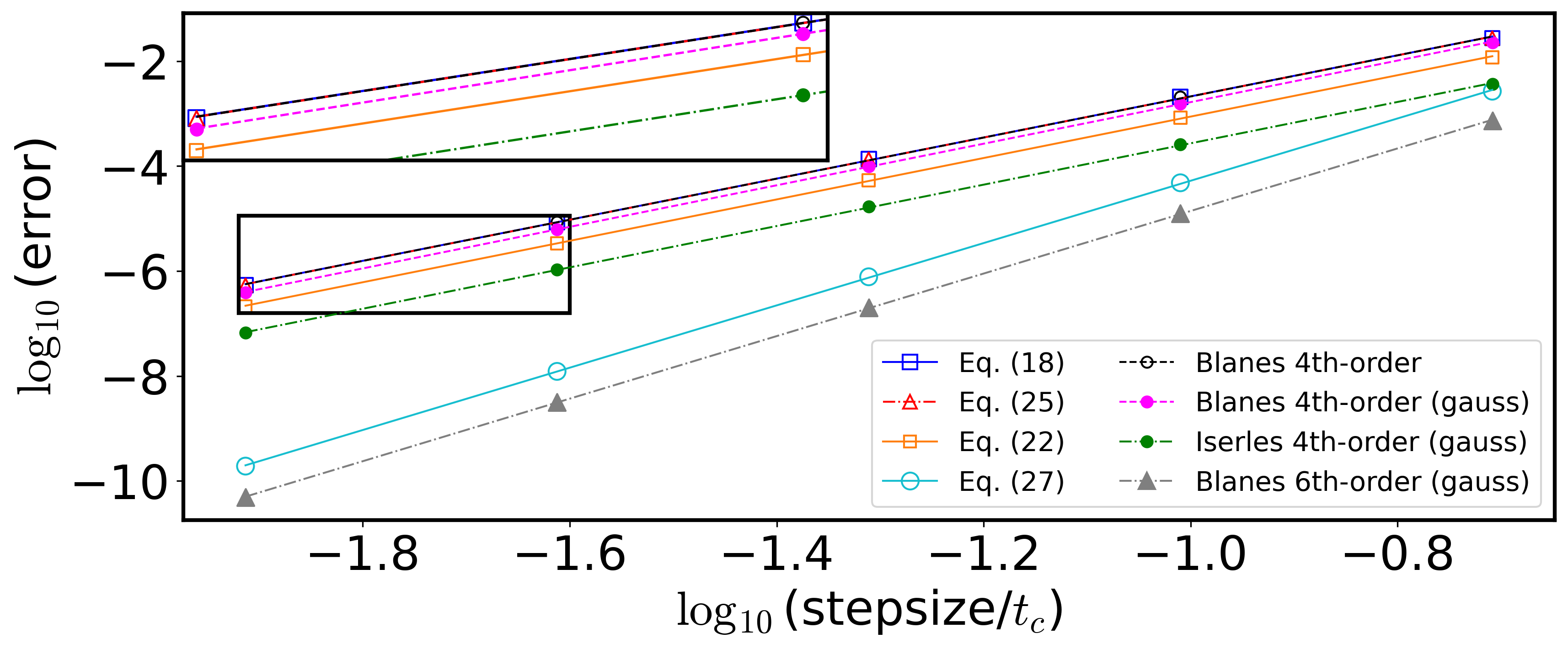} \\
(IV)\makebox[4.5in]{ }\\	\includegraphics[width=0.80\columnwidth]{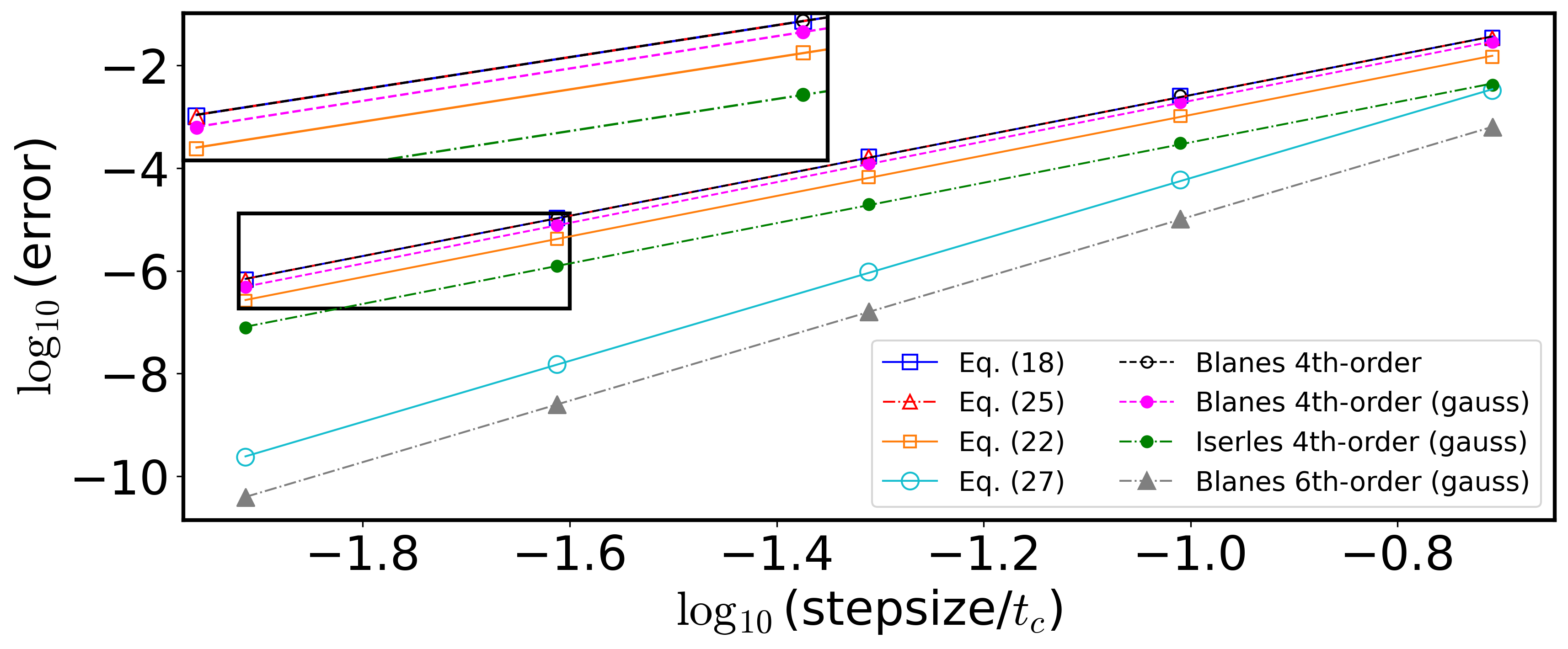} 
	\caption{Plots of $\log_{10} ({\rm error})$, where ${\rm error}$ is defined according to eq \ref{eq:error},  vs. $\log_{10} (\delta t/t_c)$, where $t_c$ is the unit time and is equal to one in the present work.  Different panels (I-IV) represent different cases of the Hamiltonian eq \ref{eq:Hamiltonian}  as listed in Table \ref{table-parameter}.  Eight different approximations are compared as listed in the legends.  Markers represent actual data points while the lines are best linear lines fitting these points.  Filled markers represent methods based on the Gaussian quadrature. Insets show close-ups of  fourth-order methods. We have also used alternating markers for some overlapping lines.}
	\label{fig:errors}
\end{figure}

\begin{table}
	\centering
	\caption{Slopes and intercepts of best linear lines fitting the data based on all methods for the case I in Figure \ref{fig:errors}.}
	\setlength{\tabcolsep}{6pt} 
	\begin{tabular}{c|c|c}
		\hline
		\hline
		\textbf{Method} & \textbf{Slope} & \textbf{y-intercept} \\
		\hline
		eq 18 & 3.996 & -1.1585  \\
		eq 25 & 3.996 & -1.1585 \\
		eq 22 & 3.999 & -0.9250  \\
		eq 27  & 5.965 & -1.7504\\
		
		Blanes 4th-order  & 3.996 & -1.1585 \\
		
		Blanes 4th-order (gauss)& 3.998 & -1.0256 \\
		Iserles 4th-order (gauss) & 3.999 & -1.0285	 \\

		Blanes 6th-order (gauss) & 5.982 & -1.6595\\
		\hline
		\hline
	\end{tabular}
	\label{table:slope_intercept_I}
\end{table}

\begin{table}
	\centering
	\caption{Slopes and intercepts of best linear lines fitting the data based on all methods for the case II in Figure \ref{fig:errors}.}
	\setlength{\tabcolsep}{6pt} 
	\begin{tabular}{c|c|c}
		\hline
		\hline
		\textbf{Method} & \textbf{Slope} & \textbf{y-intercept} \\
		\hline
		eq 18 & 3.989 & -0.6320  \\
		eq 25 & 3.995 & -0.3287 \\
		eq 22 & 3.998 & -0.5740  \\
		eq 27  & 5.990 & -0.7006\\

		Blanes 4th-order  & 3.994 & -0.3667 \\
		Blanes 4th-order (gauss)& 3.997 & -0.3193 \\
		Iserles 4th-order (gauss) & 4.000 & -0.5782	 \\

		Blanes 6th-order (gauss) & 6.000 & -0.8209\\
		\hline
		\hline
	\end{tabular}
	\label{table:slope_intercept_II}
\end{table}

\begin{table}
	\centering
	\caption{Slopes and intercepts of best linear lines fitting the data based on all methods for the case III in Figure \ref{fig:errors}.}
	\setlength{\tabcolsep}{6pt} 
	\begin{tabular}{c|c|c}
		\hline
		\hline
		\textbf{Method} & \textbf{Slope} & \textbf{y-intercept} \\
		\hline
		eq 18 & 3.917 & 1.2463  \\
		eq 25 & 3.918 & 1.2451 \\
		eq 22 & 3.947 & 0.8904 \\
		eq 27  & 5.938 & 1.6588 \\
		
		Blanes 4th-order (gauss)& 3.962 & 1.1810 \\
		
		Blanes 4th-order  & 3.918 & 1.2453 \\
		Iserles 4th-order (gauss) & 3.942 & 0.3760 \\

		Blanes 6th-order (gauss) & 5.962 & 1.1061 \\
		\hline
		\hline
	\end{tabular}
	\label{table:slope_intercept_III}
\end{table}

\begin{table}
	\centering
	\caption{Slopes and intercepts of best linear lines fitting the data based on all methods for the case IV in Figure \ref{fig:errors}.}
	\setlength{\tabcolsep}{6pt} 
	\begin{tabular}{c|c|c}
		\hline
		\hline
		\textbf{Method} & \textbf{Slope} & \textbf{y-intercept} \\
		\hline
		eq 18 & 3.919 & 1.3413  \\
		eq 25 & 3.919 & 1.3413 \\
		eq 22 & 3.948 & 0.9858 \\
		eq 27  & 5.935 & 1.7419 \\
		
		Blanes 4th-order (gauss)& 3.963 & 1.2756 \\
		
		Blanes 4th-order  & 3.919 & 1.3413 \\
		Iserles 4th-order (gauss) & 3.932 & 0.4346 \\

		Blanes 6th-order (gauss) & 5.983 & 1.0428 \\
		\hline
		\hline
	\end{tabular}
	\label{table:slope_intercept_IV}
\end{table}

Figure \ref{fig:errors} shows the errors calculated for our three fourth order expressions and the sixth order expression, and two fourth order expressions and sixth order expression by Blanes {\it et al.}\cite{blanes-bnm40} and a fourth order expression by Iserles {\it et al.}.\cite{Iserles-ptmpes357}  Markers represent actual data of calculation and lines were drawn as guide.  All the data exhibit expected scaling behavior with respect to the time step $\delta t$.  To discuss the relative efficiency of different algorithms, we also calculated average computing times in Table \ref{table:run_times}. The timing at  the largest time step is sufficient to give the general performance in all cases because of the linear scaling of ME with the number of time steps.    

For cases I and II, the difference in errors between different fourth order expressions are marginal.  The two sixth order expressions exhibit much more improved behavior and have similar accuracy.  On the other hand, for cases III and IV where there is a high frequency oscillation, appreciable differences can be seen between different approximations.  We find that our two fourth order expressions, eqs \ref{eq:u3} and \ref{eq:u4-2}, show similar performance as the 4th order expression by Blanes {\it et al.} that uses simple integration scheme.\cite{blanes-bnm40}  The 4th order expression with Gauss-Legendre quadrature by Blanes {\it et al.}\cite{blanes-bnm40} performs slightly better than the three expressions, but is worse than our 4th order expression with double commutator, eq \ref{eq:u4-1}.

\begin{table}
	\centering
	\caption{Average computing times (based on 375 runs) taken for different methods for time evolutions using 512 discrete time points.}
	\setlength{\tabcolsep}{6pt} 
	\begin{tabular}{c|c}
		\hline
		\hline
		\textbf{Method} & \textbf{Time ($s$)} \\
		\hline
		eq 18 & 0.079  \\
		eq 25 & 0.080  \\
		eq 22 & 0.087  \\
		eq 27  & 0.161\\
		
		Blanes 4th-order (gauss)& 0.080  \\
		
		Blanes 4th-order  & 0.080 \\
		Iserles 4th-order (gauss) & 0.086	 \\

		Blanes 6th-order (gauss) & 0.101 \\
		\hline
		\hline
	\end{tabular}
	\label{table:run_times}
\end{table}
Among the fourth order expressions, the one by Iserles {\it et al.},\cite{Iserles-ptmpes357} which is also  based on Gauss-Legendre quadrature,  shows the best accuracy.  However, this is the second most expensive among all the fourth order expressions since it involves evaluation at two mid points, while the evaluation at two end points of the time interval is  necessary for actual evaluation of physical observables, and since it also involves evaluation of double commutators (which is reflected in the longer times in Table \ref{table:run_times}). On the other hand, our best 4th order expression, eq \ref{eq:u4-1}, involves evaluation at only one mid point.  Thus, eq \ref{eq:u4-1} can be considered as a good choice when considering both efficiency and accuracy.  In terms of simplicity and efficiency, eq \ref{eq:u3} seems to be  the best choice.   It is interesting to note that the performance of this expression is comparable (or better than, as in the case II) to that of eq \ref{eq:u4-2}, which is based on a higher order interpolating polynomial for $M_2$.

Of the two sixth order expressions, the one by Blanes {\it et al.}\cite{blanes-bnm40}, which is based on Gauss-Legendre quadrature, clearly performs better than our sixth order expression, eq \ref{eq:u6}, for cases III and IV. This shows the importance of accurate integration. However, further improvement of eq \ref{eq:u6} is possible by incorporating similar quadrature of higher accuracy, making the latter comparable to the former. The longer times for our sixth order expression reflect the larger amount of Hamiltonian evaluations and commutators in our method. Both issues may be improved by using the fourth order polynomial approximation for all terms throughout: $ M_1^{(4)} -\frac{1}{2\hbar^2} M_2^{(4)} +\frac{i}{6\hbar^3}M_3^{(4)}  +\frac{1}{24\hbar^3}M_4^{(4)}$, although simplification of the resulting commutator expressions are non-trivial.

\section{Conclusion}
In this work, we have provided systematic approximations for the ME of the real time propagator for time dependent Hamiltonians,  based on explicit integrations of Lagrange interpolation formula.  We also tested their performance along with other existing expressions.\cite{blanes-bnm40,Iserles-ptmpes357} All of these expressions rely on evaluation of Hamiltonians at discrete values of time and are applicable to any kind of time dependent Hamiltonian that can be approximated by low order polynomials within each time interval.  We have tested these fourth order and sixth order expressions for a two-state Hamiltonian with sinusoidal time dependences given by eq \ref{eq:Hamiltonian}.  

All the plots of errors with respect to the value of time step, $\delta t$, confirm their expected order of accuracy.  Among the fourth order expressions, we find that eq \ref{eq:u3} works fairly well considering its simplicity and efficiency.  Although further test for other forms of Hamiltonian are still needed for more definite conclusion, the results provided here suggest that eq \ref{eq:u3} can serve as a useful expression for a broad range of problems involving general time dependent Hamiltonian such as in {\it ab initio} quantum dynamics\cite{castro-jcp121,goings-wcms2017} and quantum evolution of thermal ensembles.\cite{voth-jcp91,voth-acp93,jang-cen1,jang-jcp124}

All the expressions derived and tested here can be employed for numerical time evolution for a broad class of  time dependent Hamiltonians and thus can be utilized for calculating observables in time resolved spectroscopy, quantum control, and quantum sensing.  Even when the Hamiltonian is independent of time, the approximate ME expressions can be used for solving the dynamics in the interaction picture with respect to a reference Hamiltonian.  Similarly, they can also be utilized for the open system quantum dynamics when the reduced system density operator is solved in the interaction picture with respect to the zeroth order system Hamiltonian.

\acknowledgements

SJJ acknowledges major support of this research from the National Science Foundation (CHE-1900170), and a partial support in relation to quantum control of exciton dynamics from the US Department of Energy, Office of Sciences, Office of Basic Energy Sciences (DE-SC0021413).  SJJ also acknowledges support from the Korea Institute for Advanced Study (KIAS) through its KIAS Scholar program for the visit during summer. Both authors would like to thank anonymous reviewers for helpful comments and Dr. Eugene S. Mananga for helpful comments for an earlier version of this work and informing some references.   

\appendix
\section{Unitarity and symmetry of Magnus expansion}
Since the time evolution operator $U_{\delta t} (t_{k+1},t_k)$ should conserve the norm, it has to be unitary as follows:
\be
U_{\delta t}^\dagger(t_{k+1},t_{k})U_{\delta t} (t_{k+1},t_k)= U_{\delta t}(t_{k+1},t_{k})U_{\delta t}^\dagger (t_{k+1},t_k)={\rm I} , 
\ee
where ${\rm I}$ is the identity operator.    On the other hand, taking the Hermitian adjoint of eq \ref{eq:udelta-mn}, we find that
\ben
&&U_{\delta t}^\dagger(t_{k+1},t_{k}) =\exp\left \{\sum_{n=1}^\infty \frac{1}{n!} \left (\frac{i}{\hbar}\right)^n M_n^\dagger (t_{k+1},t_{k}) \right\} \nonumber \\
&&=\exp\left \{\sum_{n=1}^\infty \frac{1}{n!} \left (-\frac{i}{\hbar}\right)^n (-1)^nM_n^\dagger (t_{k+1},t_{k}) \right\}  .
 \label{eq:udelta-mn-d}
\een
Note that each term of the ME  in eq \ref{eq:udelta-mn} satisfies the following property: 
\be
M_n^\dagger(t_{k+1},t_k)=(-1)^{n-1}M_n(t_{k+1},t_k) , 
\ee
which can be verified directly for the expressions up to the fourth order shown in eqs \ref{eq:M1} - \ref{eq:M4} and be proven to be correct from the general nested commutator expression for the ME.    Therefore, eq \ref{eq:udelta-mn-d} becomes
\be
U_{\delta t}^\dagger(t_{k+1},t_{k}) =\exp\left \{-\sum_{n=1}^\infty \frac{1}{n!} \left (-\frac{i}{\hbar}\right)^n M_n(t_{k+1},t_{k}) \right\} .
\ee
Since the exponent of the above operator is the same as $U_{\delta t}(t_{k+1},t_{k})$ except for the overall minus sign, the two operators commute and their exponents can be combined, which leads to the proof of unitarity as follows: 
\ben
&&U_{\delta t}^\dagger(t_{k+1},t_{k})U_{\delta t} (t_{k+1},t_k)= U_{\delta t}(t_{k+1},t_{k})U_{\delta t}^\dagger (t_{k+1},t_k) \nonumber \\
&&=\exp\left \{\sum_{n=1}^\infty \frac{1}{n!} \left (-\frac{i}{\hbar}\right)^n \left (M_n(t_{k+1},t_{k}) -M_n(t_{k+1},t_{k}) \right)\right\} \nonumber \\
&&={\rm I} .
\een
In fact, the above identity remains true even if the Magnus expansion is truncated at finite order because the symmetry property is satisfied by each term.  This makes the finite order approximation of the Magnus expansion unitary.  

On the other hand, since the Magnus expansion does not rely on any time ordering prescription, we expect that the same expansion is valid for the backward propagation with $t_k$ and $t_{k+1}$ exchanged. 
By definition, this backward propagator satisfies the following property:
  \be
  U(t_k)=U_{\delta t} (t_k,t_{k+1})U (t_{k+1})  .
  \ee 
  This implies that  
  \ben 
  &&U_{\delta t}(t_k,t_{k+1})= \exp\left \{\sum_{n=1}^\infty \frac{1}{n!} \left (-\frac{i}{\hbar}\right)^n M_n(t_k,t_{k+1}) \right\} \nonumber \\
 && =\exp\left \{-\sum_{n=1}^\infty \frac{1}{n!} \left (-\frac{i}{\hbar}\right)^n M_n(t_{k+1},t_{k}) \right\}=U^\dagger _{\delta}(t_{k+1},t_k)  .\nonumber \\
 \een
 The above identity should remain true even if $\hbar$ is scaled by any numerical factor.   This is possible only if each term satisfies the following symmetry property:
 \be
 M_n(t_k,t_{k+1})=-M_n(t_{k+1},t_k) , \label{eq:odd-property}
 \ee 
 which can be confirmed directly for the expansions up to the fourth order, eqs \ref{eq:M1}-\ref{eq:M4}.   Let us assume that each term can be expanded with respect to $\delta t=t_{k+1}-t_k$ such that expansion coefficients are symmetric with respect to the exchange of $t_k$ and $t_{k+1}$ as follows:
 \be
 M_n (t_{k+1},t_k)=\sum_{j=0}^\infty \tilde M_n^{(j)}(t_{k+1},t_k) \delta t^j ,  \label{eq:mn-dt-exp-1}
 \ee
 where $\tilde M_n^{(j)}(t_{k+1},t_k)=\tilde M_n^{(j)}(t_k,t_{k+1})$.  In order for eq \ref{eq:odd-property} to be satisfied,  $\tilde M_n^{(j)}(t_k,t_{k+1})$ with even $j$ should be zero. 
 Therefore,  eq \ref{eq:mn-dt-exp-1} can be rewritten as 
 \be
 M_n (t_{k+1},t_k)=\sum_{j=1}^\infty \tilde M_n^{(2j-1)}(t_{k+1},t_k) \delta t^{2j-1} .  \label{eq:mn-dt-exp-2}
 \ee
 As a result, all the finite order Magnus expansions with symmetric coefficient as described above have always  even order of accuracy with respect to $\delta t$.  One simple way to achieve this property is to make Taylor series expansion of $M_n (t_{k+1},t_k)$ around $(t_{k+1}+t_k)/2$ with respect to $\delta t$.  Alternatively, our expressions in the main text show that it can also be achieved through appropriate rearranging of terms starting from Lagrange interpolation expressions. 

\section{Explicit calculation of low order ME terms}
The first order approximation for the Hamiltonian, eq \ref{eq:h1t}, can be expressed as
\be
H^{(1)}(t)= H_k+\frac{(t-t_k)}{\delta t} (H_{k+1}-H_k ) .\label{eq:linear-approx}
\ee
When this approximation is used, the integrals involved in the evaluation of $M_1^{(1)}$ and $M_2^{(1)}$ can be calculated easily as follows:
\ben
&&\int_{t_k}^{t_{k+1}} d\tau H^{(1)}(\tau)=\frac{\delta t}{2} (H_k+H_{k+1} ) ,\\
&&\int_{t_k}^{t_{k+1}} d\tau \int_{t_k}^{\tau}d\tau' [H^{(1)}(\tau), H^{(1)}(\tau')] =\frac{\delta t^2}{6} [H_{k+1},H_k] .  \label{eq:second-int}
\een
For the evaluation of integral for $M_3^{(1)}$,  we first expand the relevant Hamiltonians explicitly as follows:  
\begin{widetext}
\ben 
&& [ H^{(1)}(\tau), [H^{(1)}(\tau'),H^{(1)}(\tau'')] ] +[H^{(1)}(\tau''),[H^{(1)}(\tau'),H^{(1)}(\tau)]]  \nonumber
\\ 
&&=  [H_k+\frac{(\tau-t_k)}{\delta t} (H_{k+1}-H_k),[H_k+\frac{(\tau'-t_k)}{\delta t} (H_{k+1}-H_k ),H_k+\frac{(\tau''-t_k)}{\delta t}(H_{k+1}-H_k )] ]\nonumber \\
&&\hspace{.1in}+[H_k+\frac{(\tau''-t_k)}{\delta t} (H_{k+1}-H_k),[H_k+\frac{(\tau'-t_k)}{\delta t} (H_{k+1}-H_k ),H_k+\frac{(\tau-t_k)}{\delta t}(H_{k+1}-H_k )] ] \nonumber \\
&&  =\frac{(t_{k+1}-\tau)(\tau'-\tau'')}{\delta t^2} [H_k,[H_{k+1},H_k]] + \frac{(\tau-t_{k})(\tau'-\tau'')}{\delta t^2} [H_{k+1},[H_{k+1},H_k]]   \nonumber \\
 &&\hspace{.1in} + \frac{(t_{k+1}-\tau'')(\tau'-\tau)}{\delta t^2} [H_k,[H_{k+1},H_k]] + \frac{(\tau''-t_{k})(\tau'-\tau)}{\delta t^2} [H_{k+1},[H_{k+1},H_k]]   .\label{eq:third_order_integrand_simplified}
 \een
 Evaluation of the multiple time order integral in eq \ref{eq:M3} with the  the above integrand is now straightforward, for which we need integration of quadratic functions of time as follows:
	\ben 
	&&\int_{t_k}^{t_{k+1}} d\tau \int_{t_k}^{\tau} d\tau' \int_{t_k}^{\tau'} d\tau'' \frac{(t_{k+1}-\tau)(\tau'-\tau'')}{\delta t^2} \nonumber \\
	&&= \frac{1}{\delta t^2} \int_{t_k}^{t_{k+1}}(t_{k+1}-\tau) d\tau \int_{t_k}^{\tau} d\tau' \int^{t_k}_{\tau'} d\tau''   (\tau''-\tau') 
	\nonumber\\
	&&= \frac{1}{\delta t^2} \int_{t_k}^{t_{k+1}}(t_{k+1}-\tau) d\tau \int_{t_k}^{\tau} d\tau' \frac{(t_k-\tau')^2}{2}
	\nonumber \\
	&&= \frac{1}{6\delta t^2} \int_{t_k}^{t_{k+1}}(t_{k+1}-\tau)(\tau-t_k)^4 d\tau \nonumber \\
	&&= -\frac{1}{6\delta t^2}(\frac{1}{5}\delta t^5 - \frac{1}{4}\delta t^5) = \frac{1}{120}\delta t^3 .
	\label{eq:nested_integral_solved}
	\een
\end{widetext}	
Similarly, other three time integrations can be evaluated as follows:
\be
\int_{t_k}^{t_{k+1}} d\tau \int_{t_k}^{\tau} d\tau' \int_{t_k}^{\tau'} d\tau'' \frac{(\tau - t_k)(\tau'-\tau'')}{\delta t^2} = \frac{1}{30}\delta t^3 ,
\ee
\be
\int_{t_k}^{t_{k+1}} d\tau \int_{t_k}^{\tau} d\tau' \int_{t_k}^{\tau'} d\tau'' \frac{(t_{k+1}-\tau'')(\tau'-\tau)}{\delta t^2} = -\frac{1}{30}\delta t^3 ,
\ee
\be
\int_{t_k}^{t_{k+1}} d\tau \int_{t_k}^{\tau} d\tau' \int_{t_k}^{\tau'} d\tau'' \frac{(\tau''-t_{k+1})(\tau'-\tau)}{\delta t^2} = -\frac{1}{120}\delta t^3  .\label{eq:final_nested_integral_solved}
\ee
Combining all of them, we obtain the following simple expression:
\ben
&&\int_{t_k}^{t_{k+1}} d\tau \int_{t_k}^{\tau} d\tau' \int_{t_k}^{\tau'} d\tau'' \nonumber \\
&&\hspace{.8in}\Big( [ H^{(1)}(\tau), [H^{(1)}(\tau'),H^{(1)}(\tau'')] ] \nonumber \\
&&\hspace{.8in} +[H^{(1)}(\tau''),[H^{(1)}(\tau'),H^{(1)}(\tau)]]\Big )  \nonumber \\
&&=\frac{\delta t^3}{40}[H_{k+1}-H_k,[H_{k+1},H_k]] . \label{eq:third_order_approx}
\een

Now let us consider the quadratic approximation for the Hamiltonian, eq \ref{eq:h2t}, which can be expressed as
\ben
H^{(2)}(\tau) &=& H_k + \frac{\tau-t_k}{\delta t} \left(-3H_k + 4H_{k+\frac{1}{2}}- H_{k+1}\right) \nonumber \\
		&&+ 2\frac{(\tau-t_k)^2}{\delta t^2}\left(H_k-2H_{k+\frac{1}{2}} + H_{k+1}\right)  	.	
\een
Employing the above approximation, the integral for $M_1^{(2)}$ can be  calculated as follows: 
 \be
 \int_{t_k}^{t_{k+1}} d\tau H^{(2)}(\tau)=\frac{\delta t}{6} (H_k+4H_{k+\frac{1}{2}}+H_{k+1} ) . 
\ee

For the evaluation of $M_2^{(2)}$, let us first consider $[H^{(2)}(\tau), H^{(2)}(\tau')]$, which can be explicitly calculated and rearranged to have the following expression:
\begin{widetext}
\ben
[H^{(2)} (\tau), H^{(2)}(\tau')]&=&\left ([H_{k+1},H_k]-2[H_{k+\frac{1}{2}},H_k]\right)\left (\frac{2}{\delta t^2}(\tau-t_k)^2-\frac{2}{\delta t^2} (\tau'-t_k)^2 \right )  \nonumber \\
&&+\left ( -2[H_k,H_{k+\frac{1}{2}}]+2[H_k,H_{k+1}]-2[H_{k+\frac{1}{2}},H_k] \right )\nonumber \\
&&\hspace{.5in}\times  \frac{2}{\delta t^3} \left ( (\tau-t_k)^2 (\tau'-t_k)-(\tau-t_k)(\tau'-t_k)^2 \right) \nonumber \\
&&+\left (4[H_{k+\frac{1}{2}},H_k]-[H_{k+1},H_k]\right) \frac{1}{\delta t}\left ((\tau-t_k)-(\tau'-t_k)\right) .
\een
All the time integrations involving $\tau$ and $\tau'$ can be performed easily as follows: 
\ben
&&\frac{2}{\delta t^2}\int_{t_k}^{t_{k+1}} \int_{t_k}^\tau d\tau' \left \{ (\tau-t_k)^2-(\tau'-t_k)^2 \right \}=\frac{1}{3}\delta t^2  , \nonumber \\ \\
&&\frac{2}{\delta t^3} \int_{t_k}^{t_{k+1}} d\tau \int_{t_k}^\tau d\tau' \left \{ (\tau-t_k)^2 (\tau'-t_k)\right .  \nonumber \\
&&\left . \hspace{1in}-(\tau-t_k) (\tau'-t_k)^2 \right \} =\frac{1}{15}\delta t^2 ,\\
&&\frac{1}{\delta t}\int_{t_k}^{t_{k+1}} d\tau \int_{t_k}^\tau d\tau' \left \{ (\tau-t_k)-(\tau'-t_k)\right\} =\frac{1}{6}\delta t^2  . \nonumber \\
\een
\end{widetext}
Therefore, we find that 
\ben
&&\int_{t_k}^{t_k+1}\int_{t_k}^\tau d\tau'd\tau [H^{(2)}(\tau),H^{(2)}(\tau')]  \nonumber \\
&&=\left ([H_{k+1},H_k]-2[H_{k+\frac{1}{2}},H_k]\right) \frac{\delta t^2}{3} \nonumber \\
&&+\left (-2 [H_k,H_{k+\frac{1}{2}}]+2[H_k,H_{k+1}]-2[H_{k+\frac{1}{2}},H_{k+1}]\right)\frac{\delta t^2}{15} \nonumber \\
&&+\left (4[H_{k+\frac{1}{2}},H_k]-[H_{k+1},H_k]\right) \frac{\delta t^2}{6} \nonumber \\ 
&& =\frac{\delta t^2}{30}\left ([H_{k+1},H_k]+4[H_{k+\frac{1}{2}},H_k]+4[H_{k+1},H_{k+\frac{1}{2}}]\right)  \nonumber \\
&&=\frac{\delta t^2}{30}[H_k+4H_{k+\frac{1}{2}},H_k-H_{k+1}] . \label{eq:M_2_2 one comm}
\een


\providecommand{\latin}[1]{#1}
\makeatletter
\providecommand{\doi}
  {\begingroup\let\do\@makeother\dospecials
  \catcode`\{=1 \catcode`\}=2 \doi@aux}
\providecommand{\doi@aux}[1]{\endgroup\texttt{#1}}
\makeatother
\providecommand*\mcitethebibliography{\thebibliography}
\csname @ifundefined\endcsname{endmcitethebibliography}
  {\let\endmcitethebibliography\endthebibliography}{}

\newpage
\newpage
\begin{widetext}
\renewcommand{\theequation}{S\arabic{equation}}
\renewcommand{\thefigure}{S\arabic{figure}}
\renewcommand{\thetable}{S\arabic{table}}
\renewcommand\thepage{S\arabic{page}}
\setcounter{figure}{0}

\begin{center}
{\bf \large Supporting Information : Simple and general unitarity conserving numerical real time propagators of time dependent Schr\"{o}dinger equation based on Magnus expansion}
\end{center}
                              
\subsection{Verification of the fourth order commutator expression}
For the simplification of $M_4^{(1)}$,  we guessed a simplified form that involves an undetermined coefficient $c$ and solved the quadratic equation for $c$ such that it reproduces the original expression.  The resulting expression for the relevant integral is as follows: 
\ben
&&\int_{t_k}^{t_{k+1}} d\tau\int_{t_k}^{\tau} d\tau' \int_{t_k}^{\tau'} d\tau''\int_{t_k}^{\tau''} d\tau''' \nonumber \\ 
&&\hspace{.8in}\Big([[[H^{(1)}(\tau),H^{(1)}(\tau')],H^{(1)}(\tau'')],H^{(1)}(\tau''')] \nonumber \\ 
&&\hspace{.8in}+[H^{(1)}(\tau),[[H^{(1)}(\tau'),H^{(1)}(\tau'')],H^{(1)}(\tau''')]] \nonumber \\
&&\hspace{.8in}+[H^{(1)}(\tau),[H^{(1)}(\tau'),[H^{(1)}(\tau''),H^{(1)}(\tau''')]]] \nonumber \\
&&\hspace{.8in}+[H^{(1)}(\tau'),[H^{(1)}(\tau''),[H^{(1)}(\tau'''),H^{(1)}(\tau)]]]\Big) \nonumber \\
&&= \frac{\delta t^4}{420}[\frac{1}{c}H_k-H_{k+1},[H_{k+1} - cH_k,[H_{k+1},H_k]]] , 
\label{eq:fourth_order_final_expression}
\een
where $c = -\frac{5 - \sqrt{21}}{2}$.  Note that the choice of $c$ is not unique because the algebraic equation under-determined.  We also verified this expression using a computer algebra system known as SymPy.\cite{meurer-pcs3} 

\label{sec:verification_of_fourth_order}
The steps to verify Eq. (\ref{eq:fourth_order_final_expression}) are as follows:

1) Importing sympy and declaring the variables found in Eq. (\ref{eq:fourth_order_final_expression}). 
\begin{lstlisting}[language=Python]
import sympy as sp

H = sp.Function('H',commutative=False)
t, t_0 = sp.symbols("t t_0",constant=True)
tau_1, tau_2, tau_3, tau_4 = sp.symbols('tau_1 tau_2 tau_3 tau_4')
\end{lstlisting}
2) Coding the commutators and integrals on the LHS of Eq.(\ref{eq:fourth_order_final_expression}) as follows:
\begin{lstlisting}
from sympy.physics.quantum import Commutator
int_1 = 0
int_1 += Commutator(Commutator(Commutator(H(tau_1),H(tau_2)),H(tau_3)),H(tau_4))
int_1 += Commutator(H(tau_1),Commutator(Commutator(H(tau_2),H(tau_3)),H(tau_4)))
int_1 += Commutator(H(tau_1),Commutator(H(tau_2),Commutator(H(tau_3),H(tau_4))))
int_1 += Commutator(H(tau_2),Commutator(H(tau_3),Commutator(H(tau_4),H(tau_1))))
int_1 *= sp.Rational(1,12)
integral = sp.Integral(int_1,(tau_4,t_0,tau_3))
integral= sp.Integral(integral,(tau_3,t_0,tau_2))
integral = sp.Integral(integral,(tau_2,t_0,tau_1))
integral= sp.Integral(integral,(tau_1,t_0,t))
\end{lstlisting}
3) Define the linearization function equivalent to Eq. (63) in the main text, apply it to the integral, and simplify the expression:
\begin{lstlisting}
def linearization(tau):
return H(t_0) + ((tau-t_0)/(t-t_0) * (H(t)-H(t_0)))
result = integral.replace(lambda x:x.func==H, lambda x:linearization((x.args[0])),simultaneous=True)
result = result.expand().doit(manual=True).simplify()
result = result.subs(t_0,0,simultaneous=True).simplify()
\end{lstlisting}
4) Code the RHS of Eq. (\ref{eq:fourth_order_final_expression}) and test for equality. In SymPy, equality testing is done by subtracting the two expressions and simplifying. The last line of code prints this difference, which will equal zero for equivalent expressions.
\begin{lstlisting}
c = sp.symbols('c')
RHS = sp.expand(Commutator(-1/c*H(0)+1*H(t),Commutator(c*H(0)+-1*H(t),
Commutator(H(t),H(0)))))/5040*t**4
RHS = RHS.subs(c, sp.Rational(-5,2) - sp.sqrt(21)/2).doit()
#RHS = RHS.subs(c,sp.Rational(-5,2) + sp.sqrt(21)/2).doit() #alternative value for c
print((result - RHS).simplify())
\end{lstlisting}

\subsection{Populations with high frequency terms removed}
See Fig. S1.

\begin{figure}		
	(III)\makebox[4.5in]{}\\ \includegraphics[width=0.50\columnwidth]{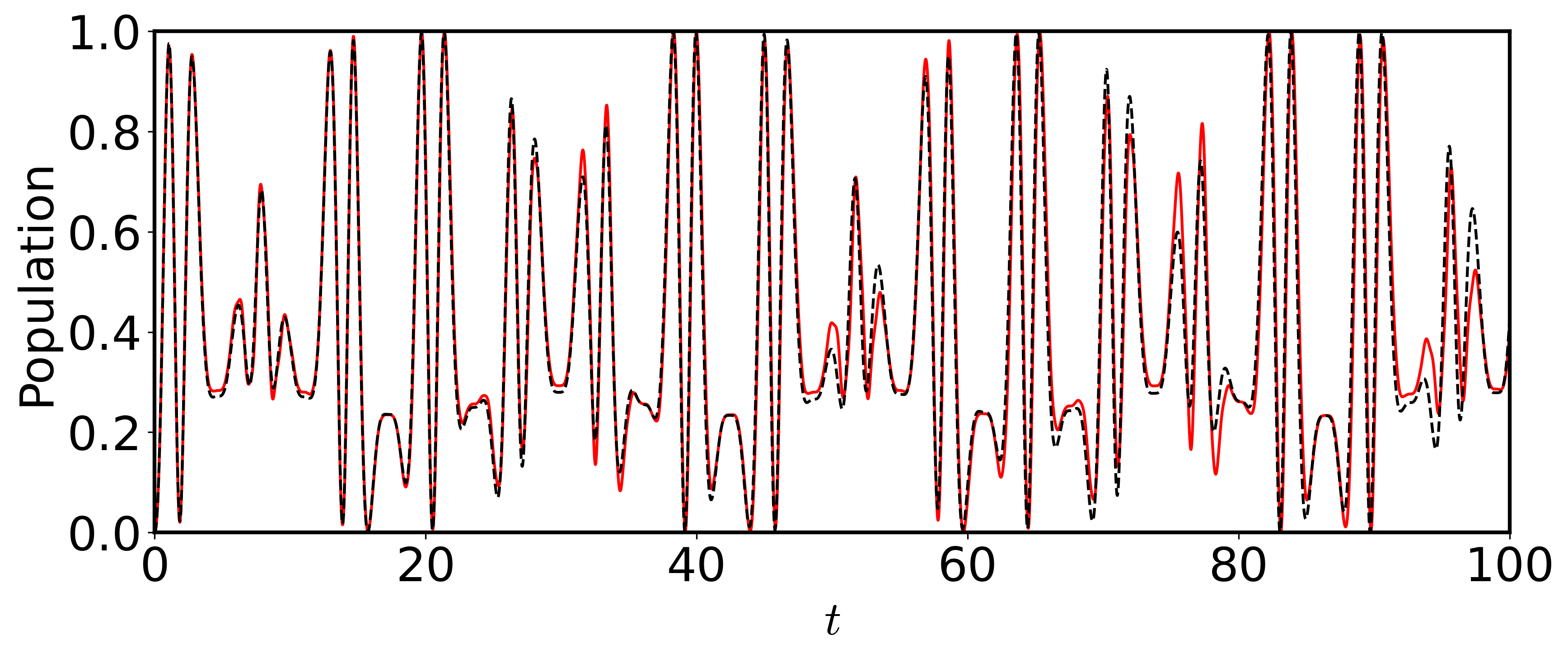} \\
	(IV)\makebox[4.5in]{}\\	\includegraphics[width=0.50\columnwidth]{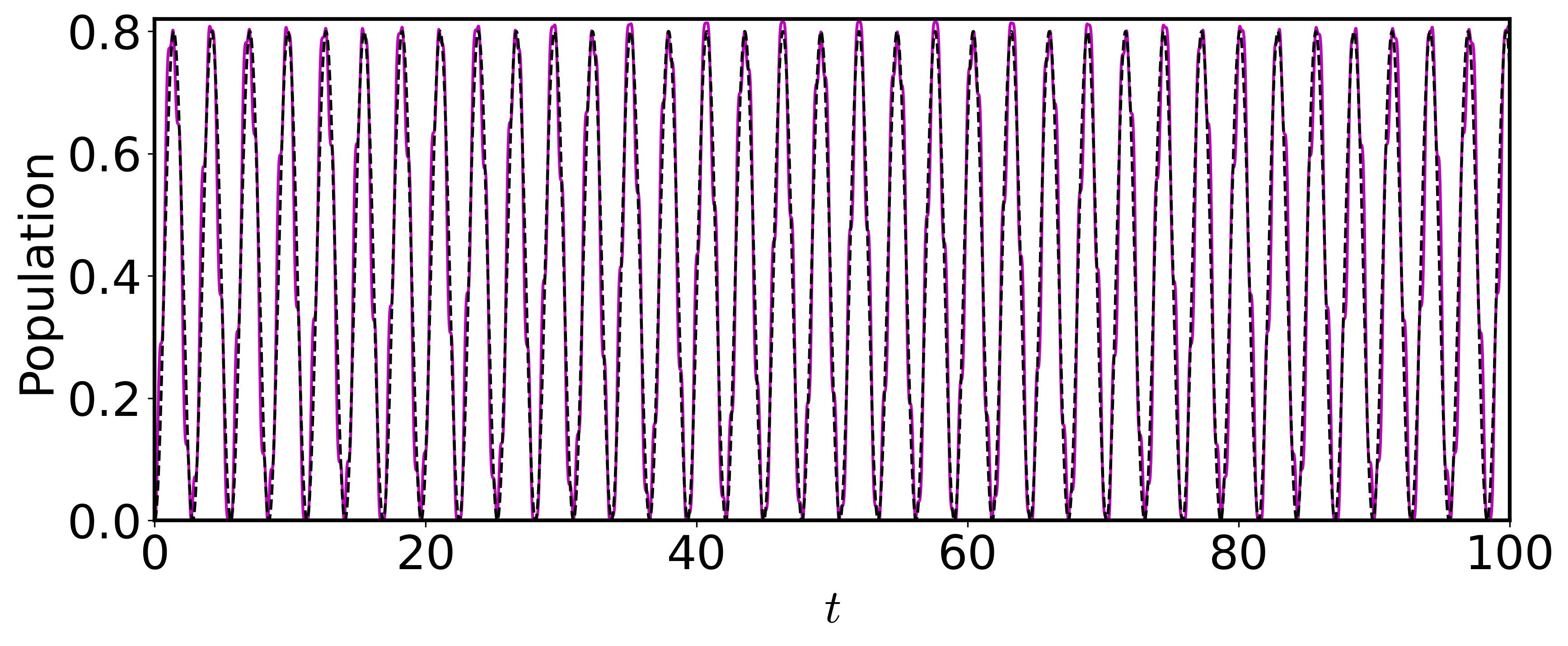} \\
	\caption{Populations of  the excited state $|1\rangle$ for high frequency cases III and IV with parameters listed in Table 1 of the main text. At time zero, all the populations are at the state $|0\rangle$. Black dashed lines represent populations with the high frequency term removed. }
	\label{fig:populations}
\end{figure}

\end{widetext}
\providecommand{\latin}[1]{#1}
\makeatletter
\providecommand{\doi}
  {\begingroup\let\do\@makeother\dospecials
  \catcode`\{=1 \catcode`\}=2 \doi@aux}
\providecommand{\doi@aux}[1]{\endgroup\texttt{#1}}
\makeatother
\providecommand*\mcitethebibliography{\thebibliography}
\csname @ifundefined\endcsname{endmcitethebibliography}
  {\let\endmcitethebibliography\endthebibliography}{}


\end{document}